\documentclass[fleqn,usenatbib]{mnras}

\usepackage[T1]{fontenc}

\DeclareRobustCommand{\VAN}[3]{#2}
\let\VANthebibliography\thebibliography
\def\thebibliography{\DeclareRobustCommand{\VAN}[3]{##3}\VANthebibliography}

\usepackage{graphicx}	
\usepackage{amsmath}	
\usepackage{amssymb}	
\usepackage{booktabs}
\usepackage{newtxtext,newtxmath}
\usepackage{multirow}
\usepackage{lscape}

\newcommand\totaldetections{417}
\newcommand\totalstars{63}
\newcommand\farawayobjects{2}
\newcommand\sourcesremaining{352}
\newcommand\sfsources{83}
\newcommand\diffusesources{269}

\newcommand\outerexcess{0.065}
\newcommand\outermedian{64}

\newcommand\outerspread{25--344}
\newcommand\innerexcess{0.241}
\newcommand\innermedian{48}

\newcommand\innerspread{4--277}

\title[UVIT view of Cen A]{UVIT view of Centaurus A; a detailed study on positive AGN feedback}

\author[P. Joseph et al.]{
Prajwel Joseph,$^{1,2}$\thanks{E-mail: prajwel.pj@gmail.com (PJ)}
P. Sreekumar,$^{3,2}$
C. S. Stalin,$^{2}$
K. T. Paul,$^{1}$
Chayan Mondal,$^{4}$
Koshy George$^{5}$ \newauthor
\ and Blesson Mathew$^{1}$
\\
$^{1}$Department of Physics, CHRIST (Deemed to be University), Bangalore 560029, India\\
$^{2}$Indian Institute of Astrophysics, Bangalore 560034, India\\
$^{3}$ISRO HQ, Antariksh Bhavan, Bangalore 560094, India\\
$^{4}$ Inter-University Center for Astronomy and Astrophysics, Pune 411007, India\\
$^{5}$Faculty of Physics, Ludwig-Maximilians-Universität, Scheinerstr. 1, Munich 81679, Germany
}

\date{Accepted XXX. Received YYY; in original form ZZZ}

\pubyear{2021}

\begin{document}
\label{firstpage}
\pagerange{\pageref{firstpage}--\pageref{lastpage}}
\maketitle

\begin{abstract}
Supermassive black holes at the centre of active 
galactic nuclei (AGN) produce relativistic jets 
that can affect the star formation characteristics 
of the AGN hosts. 
Observations in the ultraviolet (UV) band can 
provide an excellent view of the effect of 
AGN jets on star formation.
Here, we present a census of star formation 
properties in the Northern Star-forming Region (NSR) 
that spans about 20 kpc of the large radio source 
Centaurus A hosted by the giant elliptical 
galaxy NGC 5128.
In this region, we identified \sourcesremaining{} 
UV sources associated with Cen A using new 
observations at an angular resolution of 
$<$1.5 arcseconds observed with the 
Ultra-Violet Imaging Telescope (UVIT) onboard AstroSat. 
These observations were carried out in one 
far-ultraviolet (FUV; $\lambda_{\text{mean}}$ = 1481 \AA) and 
three near-ultraviolet (NUV; with $\lambda_{\text{mean}}$ of 
2196 \AA, 2447 \AA, and 2792 \AA, respectively) bands. 
The star-forming sources identified in UV tend to lie in 
the direction of the jet of Cen A, thereby suggesting jet 
triggering of star formation. 
Separating the NSR into Outer and Inner regions, 
we found the stars in the Inner region to have a 
relatively younger age than the Outer region, 
suggesting that the two regions may have different 
star formation histories.
We also provide the UVIT source catalogue in 
the NSR.

\end{abstract}




\section{Introduction}

Centaurus A (Cen A) is a large radio source occupying 
8$\degr$ $\times$ 4$\degr$ in the sky and associated 
with the early-type galaxy NGC 5128 
\citep{israel1998centaurus}. 
\cite{bolton1949positions}, based on radio 
observations of Cen A, was the first to note 
its association with NGC 5128.
The radio emission is from the active 
galactic nucleus (AGN) at the central 
region of NGC 5128.  
Cen A has been observed in a wide range 
of wavelengths and studied in detail due to  
its proximity of 3.8 Mpc \citep{harris2010}. 
These studies reveal the presence of a warped 
disk and optical shells with associated neutral 
hydrogen and CO emission; they are believed to be 
the aftereffects of a past merger activity
\citep{quillen1993warpeddisk, 
       malin1983shell, 
       schiminovich1994neutralhydrogen, 
       charmandaris2000moleculargas}. 
According to a recent study using simulations 
of Cen A, the merger occurred $\sim$2 Gyr ago 
\citep{wang2020mergerage}. 
       
Post-merger galaxies show an enhancement in the
neutral and molecular gas content compared to galaxies 
of similar stellar mass with no recent merger activity
\citep{ellison2018enhancedNeutralgas, 
       pan2018enhanceMolecularGas}.
Simulations suggest that the aforementioned 
enhancement takes place through the depletion 
of ionised gas \citep{moreno2019mergersimulation}. 
Mergers are also known to trigger AGN activity 
in the local universe \citep{ellison2019mergertriggerAGN}. 
Due to proximity, signatures of a recent merger, 
and the presence of an AGN, Cen A allows us to study 
the influence of AGN and merger activity on star 
formation in great detail.
In Cen A, the age of observed radio structure is
estimated to be of the order of a Gyr, and there could 
be a possibility of the AGN activity being triggered 
by the recent merger \citep{eilek2014dynamicmodelRadio}.

While a merger can trigger AGN activity, it may 
also produce tidal structures which should disappear 
after a few Gyrs. 
Neutral and molecular gases are known to be associated
with tidal structures \citep{duc2013tidesbook}.
There exist few observed cases where AGN interacts
with the gas content located outside the galaxy to 
induce star formation; this is an effect known as 
positive AGN feedback
\citep{van1985minkowski, 
       van1993induced3c285, 
       bicknell2000jet4c41, 
       ngc1275canning2014filamentary}.
The presence of AGN can also inhibit star formation 
in the galaxy, and therefore the effect of AGN feedback 
is complex.
Cen A is believed to have experienced positive feedback
where the jet launched from the central black hole 
might have interacted with the gas present around 
the galaxy retained from the past merger activity. 
This jet ploughing through the cold (HI \& H$_2$) gas 
could be the trigger for star formation. 
One of the pioneering studies in this regard was 
carried out by \cite{mould2000jetCenA}, where they 
discussed the observed OB associations as arising 
due to jet interaction with the gas cloud. 
\cite{neff2015complex} argue that a galactic wind
could also be at play along with the jet.
\cite{oosterloo2005anomalous} proposed that the jet 
interacted with a large neutral hydrogen cloud 
present on the northeast side of Cen A based 
on kinematic information.

An aspect that has been crucial in Cen A AGN-cloud 
interaction studies is the complex morphology 
of AGN emission revealed by radio/X-ray 
observations 
\citep{feigelson1981xray, 
            junkes1993radio,
            dobereiner1996rosat,
            kraft2000chandra,
            kraft2002chandra,
            hardcastle2003radio}.
The observed radio emission at scales of giant, 
medium and small level radio lobes has been 
classified in the literature and studied.  
From the centre of NGC 5128, at about 15 kpc 
north-east (NE), lies the northern middle lobe (NML). 
The NML appears to be an extension of the 
northern inner lobe (NIL) seen at a distance of 
about 8 kpc from the centre of NGC 5128; however, 
the connection between them seems unclear 
\citep{neff2015radio, mckinley2018jet}.
The origin of the NML is also of considerable 
interest and there exist different explanations 
in the literature 
\citep{morganti1999centaurus,
      saxton2001centaurus, 
      kraft2009jet, 
      2010gopal, 
      neff2015complex}.

Available observations point to the presence of 
young massive stars in the region, which could 
be induced by the relativistic jet of Cen A 
\citep{rejkuba2002radio}. 
Ultraviolet (UV) observations using GALEX by 
\cite{neff2015complex} too have shown the presence 
of UV bright filaments with young star clusters
in a section spanning 20 kpc northward; 
in the literature, they are referred to as 
Inner filament, seen at a distance of $\sim$8 kpc 
from the centre of NGC 5128 and near NIL, and Outer 
filament at $\sim$15 kpc and inside the NML.
Collectively, we call this northward section from 
NGC 5128 as the northern star-forming region (NSR). 
Among the various tracers to understand the 
nature of stars in regions with star formation, 
UV is particularly important due to its 
sensitivity towards stars formed over the past 
10--200 Myr \citep{kennicutt2012star}. 
It is also possible to constrain the ages of 
stellar populations up to several hundreds of Myrs 
with UV colours \citep{bianchi2011galex}. 
Available GALEX observations are of low resolution, 
and therefore high-resolution observations in the 
UV band would reveal a wealth of information on the 
nature of the stellar population in the NSR, 
as well as, provide clues to the jet induced 
star formation activity. 

We performed UV observations of the 
NSR using Ultra-Violet Imaging Telescope (UVIT) 
in different UV filters with an angular resolution 
better than 1.5 arcseconds. 
This work presents the results from our 
new observations in the UV band. 
Our paper is organised as follows. 
In Section 2, we describe the observations, 
reduction procedures, and detection of UV sources; 
Section 3 describes the characterisation of 
the UV sources,
and the results are discussed in the final sections. 
We adopt a flat Universe cosmology with 
H$_{\rm{o}} = 71\, \mathrm{km\, 
s^{-1}\, Mpc^{-1}}$, $\Omega_{\rm{M}} = 0.27$, 
$\Omega_{\Lambda} = 0.73$ \citep{komatsu2009five}. 
Cen A resides at a distance of 3.8 Mpc, and one 
arcsecond corresponds to $\sim$18 pc at this 
distance \citep{harris2010}.


\section{Observations \& Analysis}

Our observations of Cen A NSR were made using the 
UVIT telescope.
UVIT is one of the payloads aboard AstroSat with 
observing capabilities in FUV and NUV channels 
\citep{tandon2017firstresults, ghosh2021orbit}.
In each channel, there are filters available
to probe specific bands in the respective 
wavelength ranges. 
Both FUV and NUV telescopes have a field of view 
of $\sim$28 arcminutes and an angular resolution 
of $<$1.5 arcseconds (FWHM).
The FUV and NUV observations of NSR were made with 
a UVIT pointing centred on RA = 201$\degr$.602291,
Dec = -42$\degr$.844600. 
Our observations of NSR were carried out from 
2018-03-14 to 2018-03-15 (UT) with a total 
scheduled exposure of $\sim$24000 seconds in both 
the channels, which took 17 AstroSat orbits.
We did imaging in broadband FUV, narrowband NUV, 
and broadband NUV filters with varying integration times.
Although N245M filter exposure time is $\sim$5 times less than 
N219M filter, N245M is $\sim$5 times more sensitive than
N219M \citep[see][table 3]{tandon2020additional}.
Table~\ref{tab:obs_summary} gives details on UVIT 
FUV and NUV imaging.

The science ready UVIT images were downloaded 
from the ISSDC AstroSat archive.
These images were reduced using the UVIT Level2 
pipeline version 6.3 \citep{ghosh2021performance, 
ghosh2022automated} by the UVIT Payload Operations centre 
and made available to ISSDC.
The pipeline corrected for the telescope drift 
of UV data in each orbit using the VIS channel 
and then aligned and combined the orbit-wise images.
These data were processed using the calibration database 
version 20190625, which contains the latest flat fielding and 
distortion correction information.
The astrometric calibration of the images 
was carried out using the Astrometry.net 
package \citep{lang2010astrometry}.
All NUV images were re-projected to the 
FUV image projection using the 
flux-conserving spherical polygon 
intersection method of reproject package 
\citep{robitaille2020reproject}. 
The final images have a plate scale of 
0.4169 arcseconds per pixel.


\begin{table*}
\centering
\caption{Summary of UVIT Level2 data. 
$\lambda_{\text{mean} }$ and $\Delta\lambda$ 
values are taken from \protect\cite{tandon2020additional}.
The quoted exposure times are from the 
science-ready images.
For a signal-to-noise ratio (SNR) of 3, 
we estimated the corresponding AB magnitude values
for UVIT observations in each filter (given in
the 6th column). 
The corresponding flux values are also provided.}
\label{tab:obs_summary}
\begin{tabular}{@{}llrrrrr@{}}
\toprule
\multicolumn{1}{c}{Channel} &
  \multicolumn{1}{c}{Filter} &
  \multicolumn{1}{c}{$\lambda_{\text{mean}}$ (\AA)} &
  \multicolumn{1}{c}{$\Delta\lambda$ (\AA)} &
  \multicolumn{1}{c}{Exposure time (s)} &
  \multicolumn{1}{c}{AB mag$_{(\text{SNR}=3)}$} &
  Flux$_{(\text{SNR}=3)}$ (erg cm$^{-2}$ s$^{-1}$ \AA$^{-1}$) \\ \midrule
FUV & F148W & 1481 & 500 & 23677 & 26.24 & 1.588E-18 \\
NUV & N219M & 2196 & 270 & 10588 & 24.12 & 5.094E-18 \\
NUV & N245M & 2447 & 280 & 2148  & 24.75 & 2.281E-18 \\
NUV & N279N & 2792 & 90  & 7473  & 23.55 & 5.298E-18 \\ \bottomrule
\end{tabular}
\end{table*}

\subsection{Cen A: NUV and FUV imaging}

\begin{figure*}
	\includegraphics[width=\textwidth]{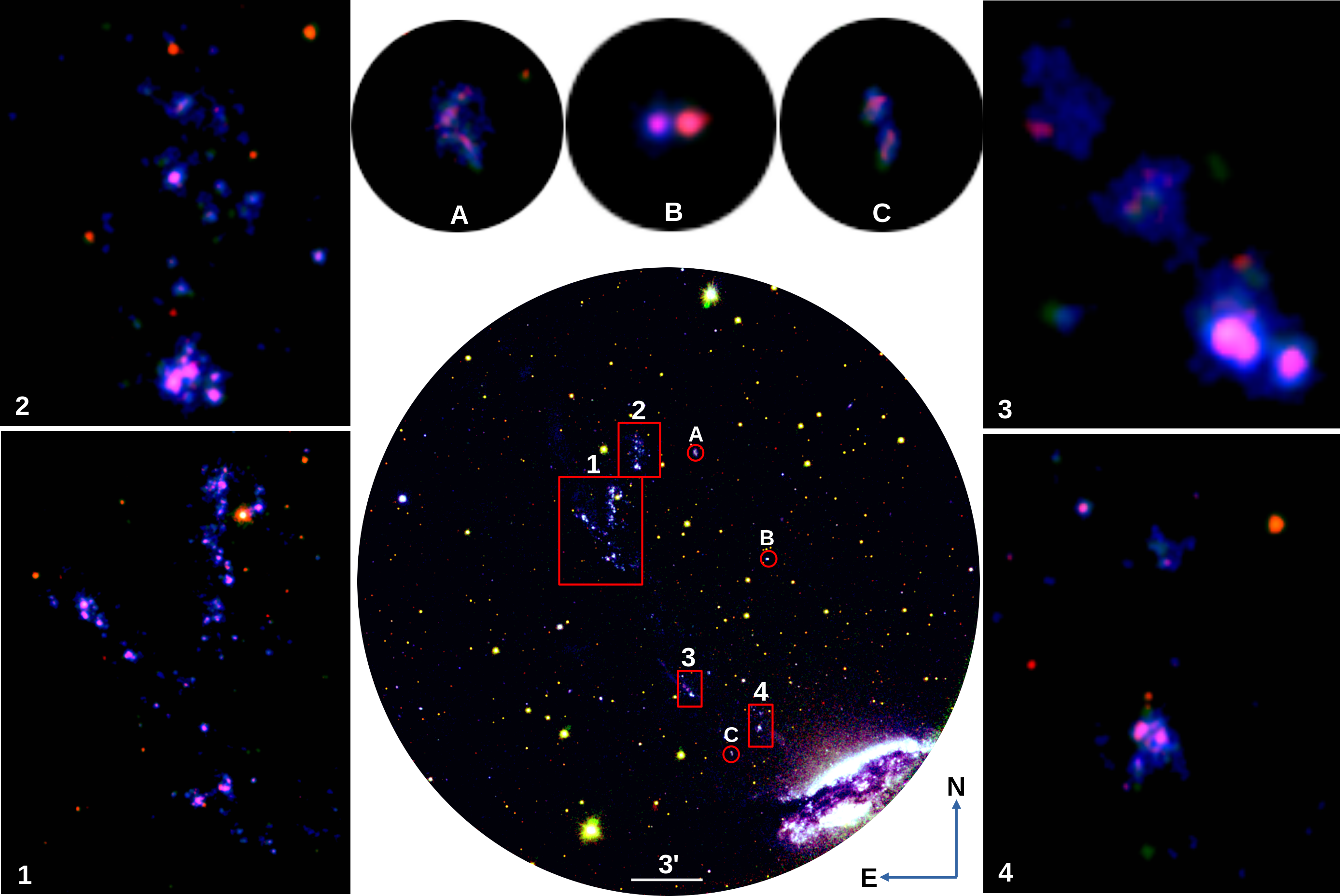}
    \caption{UVIT image of Cen A with F148W filter in blue,
             N219M in green, and N245M in red colours.
             The image has been processed to make the
             features stand out. 
             The central large circular image shows the
             $\sim$0.5$\degr$ UVIT field of view.
             The white horizontal line is shown for 
             scale and has a length of 3 arcminutes.
             Thumbnails placed around the main image are 
             zoomed sections of the image chosen to
             showcase the imaging capabilities of UVIT.
             Here, Thumbnails 1 and 2 are part of the Outer filament, 
             Thumbnail 3 shows part of the Inner filament, and
             Thumbnail 4 shows new candidates of jet induced 
             star-forming sources.
             Circular thumbnails A-C highlights a few of the interesting 
             sources found in the image.
             Circular thumbnails A and C show UV sources near the Outer 
             and Inner filaments, respectively.
             Circular thumbnail B has a pair of stars, with one on the 
             left brighter in FUV than the right one.
             The source on the right in Circular thumbnail B 
             is a RR Lyr type variable star 
             (J132546.5–425141.3 in \protect\citealt{kinman2014rrlyr}).
             }
    \label{fig:collage}
\end{figure*}

A colour composite UVIT image with a circular size of 
$\sim$0.5$\degr$ diameters created from 
F148W, N219M and N245M passbands is shown 
in Fig.~\ref{fig:collage}. 
NGC 5128 can be seen at the bottom right part 
of the image. 
The galaxy has a prominent UV bright disk with a 
dark band; this morphology results from 
a strongly warped disk rich in dust and star formation
\citep{quillen2006spitzer}.
UV bright star-forming structures seen outside 
the galaxy disk and in the NSR are marked in 
red boxes in Fig.~\ref{fig:collage}.



Compared to previous GALEX observations, 
the higher spatial resolution of UVIT reveals 
the intricate structure of the star-forming 
regions in more detail. 
Note that N219M-N245M medium band filter combination 
in the NUV channel became helpful in correcting for 
the dust attenuation (see appendix \ref{sec:colour_excess}). 
The following sections present a detailed 
analysis of the star-forming sources revealed 
by the new UVIT observations.

\begin{figure}
	\includegraphics[width=\columnwidth]{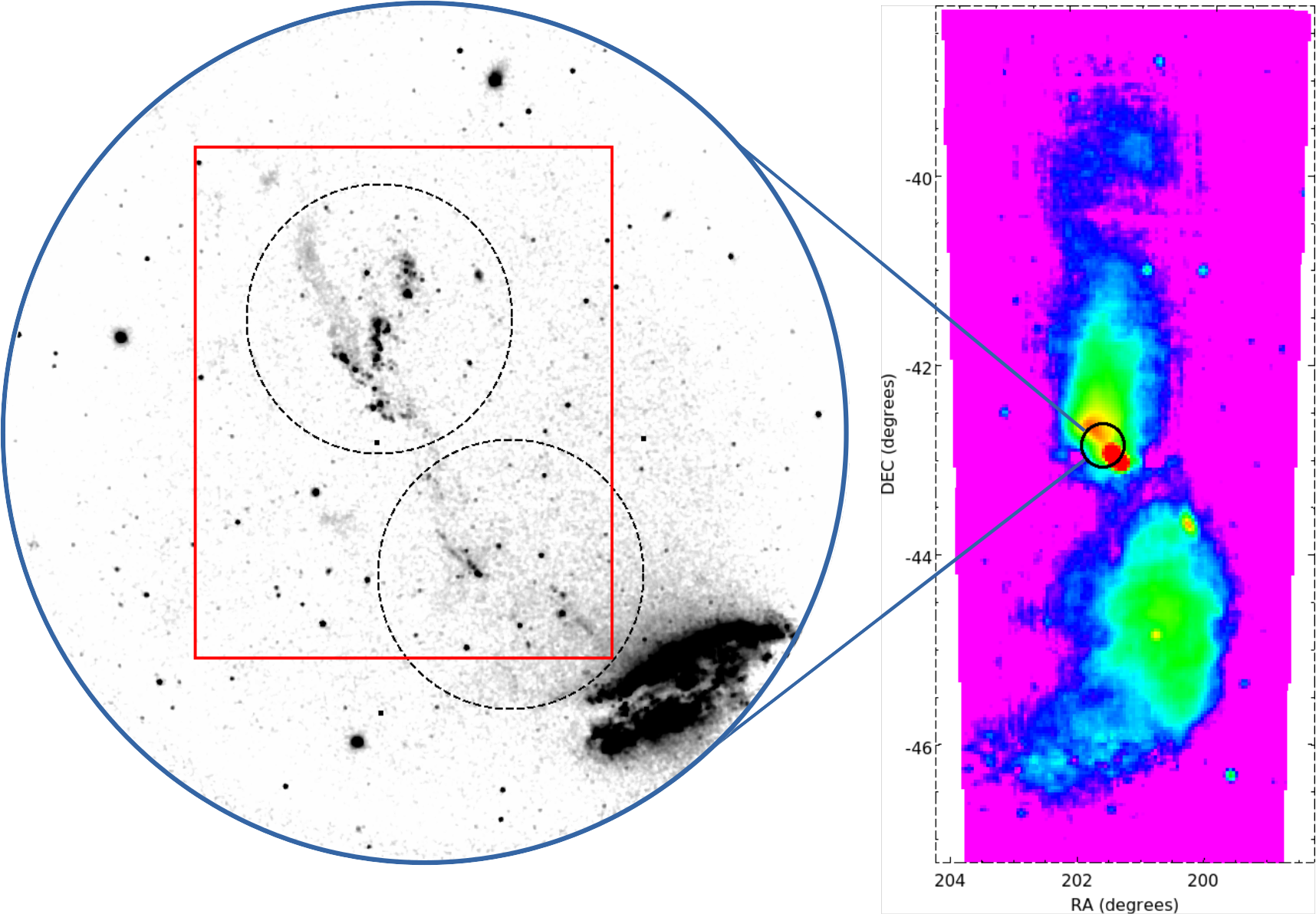}
    \caption{\textit{Left}: F148W filter image of Cen A.
             The image has been processed to make the
             faint and diffuse features visible. 
             The red box depicts the area considered 
             for analysis in this paper.
             Two dashed circles represent the 
             Outer region (see Fig.~\ref{fig:outer_region})
             and the Inner region (see 
             Fig.~\ref{fig:inner_region}).
             \textit{Right}: 6.3 cm total intensity map of Cen A 
             from \protect\cite{junkes1993radio}.
             The black circle of 28 arcminute diameter is
             the field of view of our UVIT observations.
             }
    \label{fig:showcase}
\end{figure}



\subsection{Detection of objects}

\begin{figure}
	\includegraphics[width=\columnwidth]{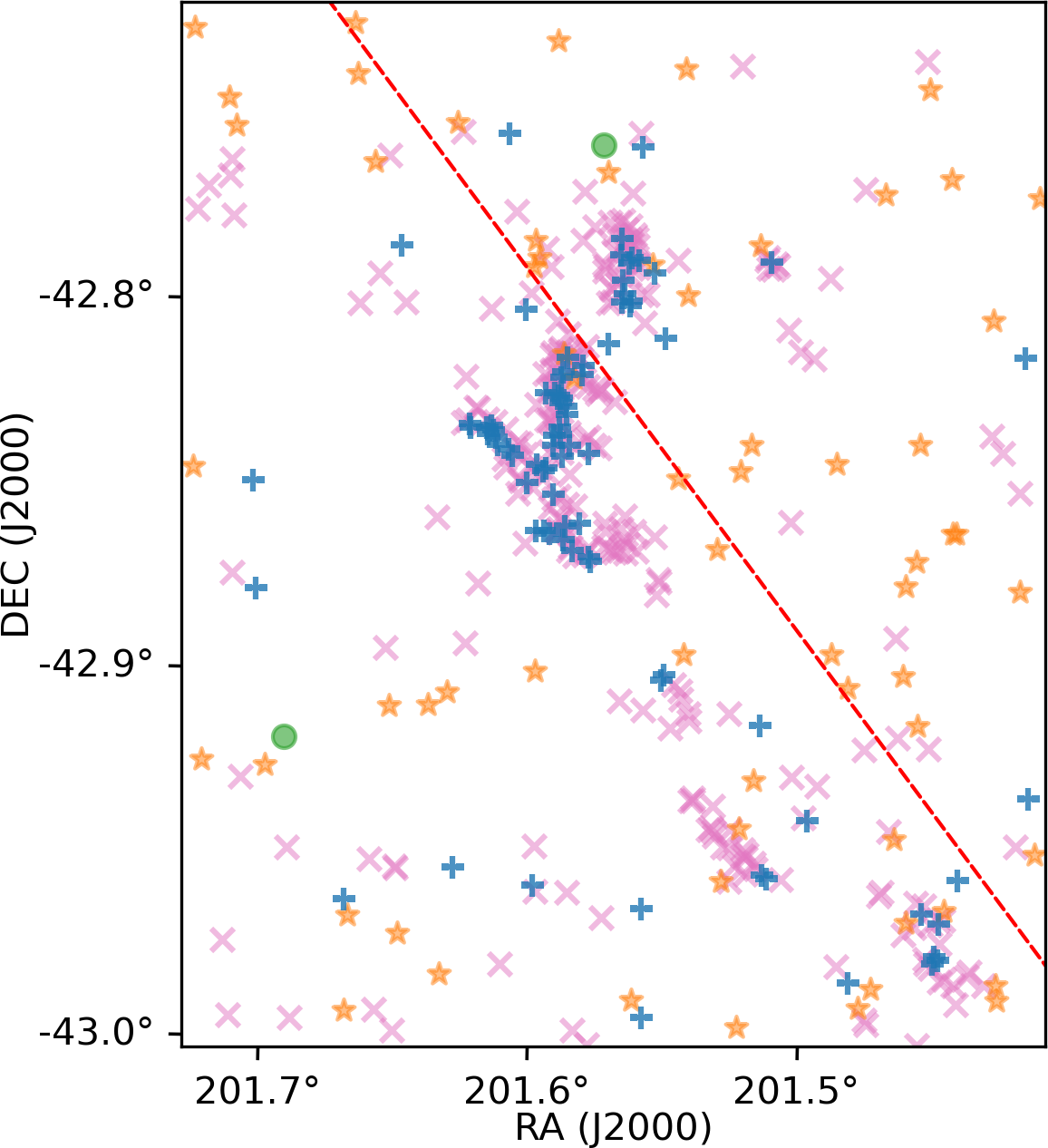}
    \caption{The \totaldetections{} sources detected in the 
             NSR (red box in Fig.~\ref{fig:showcase}) are plotted. 
             The orange stars are the \totalstars{} galactic sources, 
             green circles are \farawayobjects{} distant galaxies,
             blue crosses are \sfsources{} star-forming sources,
             and \diffusesources{} diffuse sources are depicted
             by pink X shaped markers.
             The red dashed line is the vector path through NSR
             representing the potential path associated
             with a past jet activity (defined in 
             section~\ref{sec:outer_region}).}
    \label{fig:sources}
\end{figure}

\begin{figure}
	\includegraphics[width=\columnwidth]{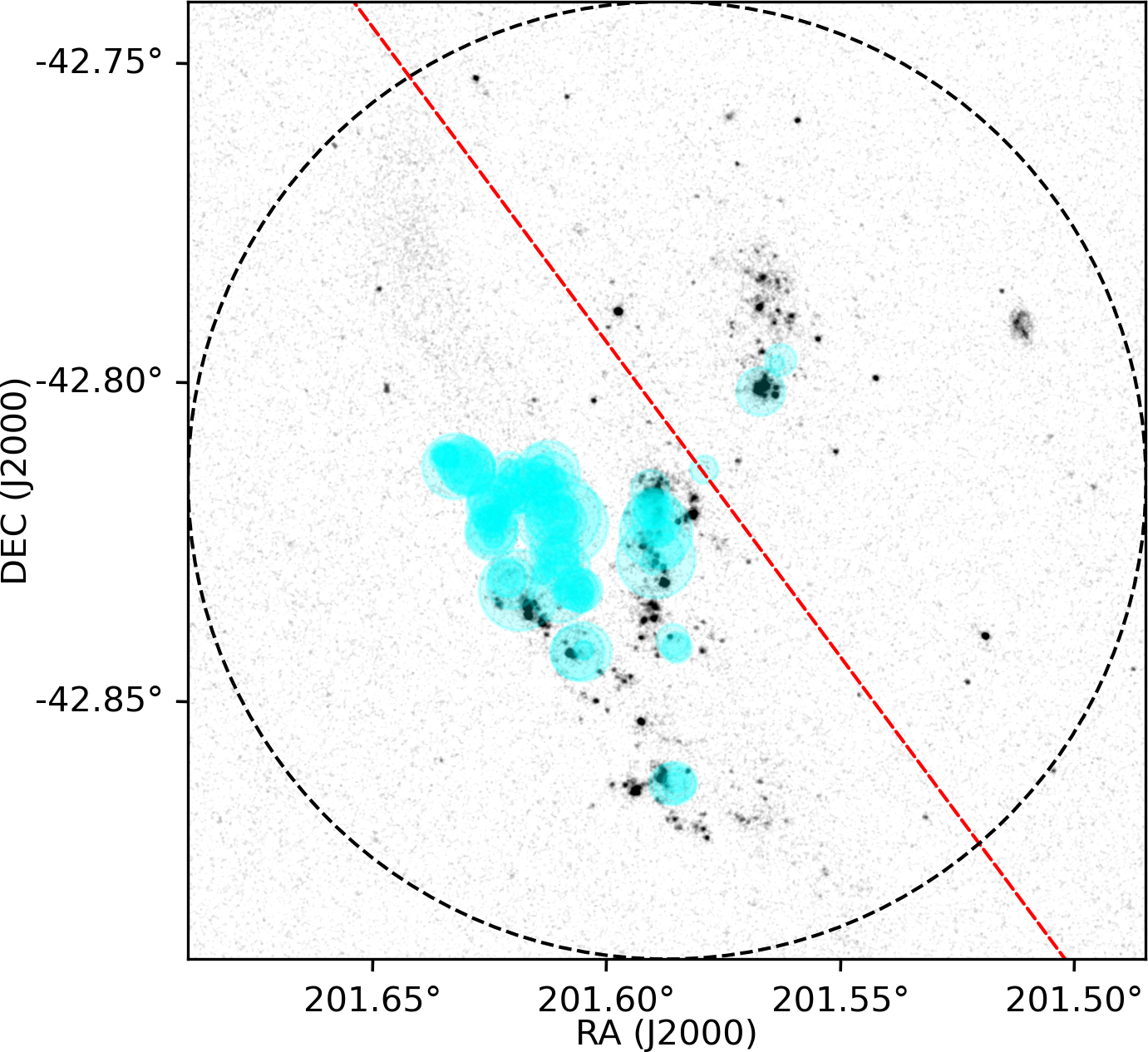}
    \caption{A closer look at the V-shaped Outer filament using
             the F148W image.
             The red dashed line is the vector path.
             The black dashed circle has a radius of 4.5 arcminutes
             with a centre at RA = 201$\degr$.5884734,
             Dec = -42$\degr$.8144597. 
             The semi-transparent cyan circles 
             represent the molecular clouds, 
             with the radius proportional to the cloud mass. 
             The molecular cloud data used here is from ALMA 
             observations by \protect\cite{salome2017inefficient}.}
    \label{fig:outer_region}
\end{figure}

Our study concentrates on recent star formation
observed near and inside the Inner and 
Outer filaments in the NSR. 
An area encompassing these filaments and their 
surroundings were chosen for further analysis. 
New UVIT observations reveal UV bright sources 
in the NSR shown as a red box in Fig.~\ref{fig:showcase}.
The red box has a centre at 
RA = 201$\degr$.570662, DEC = -42$\degr$.860251
and a size of 17$\times$14 arcminutes.
Most of the UV bright sources are seen near 
the Outer filament. 
Here, UV sources are roughly segregated into a 
double structure with a gap in an otherwise 
continuous distribution of sources 
(see figures~\ref{fig:sources} \& \ref{fig:outer_region}). 
This double structure consists of a V-shaped 
segment and a smaller elongated segment.
When it comes to the Inner filament, UV sources
are present, but not as many as in the Outer filament.
They are observed as a single elongated structure. 
Between the Inner filament and NGC 5128 lies another 
set of UV sources with a sparse extent 
similar to the Inner filament. 
To study such a diverse collection of sources,
we attempted to detect all sources in the
field, find properties, and classify them.  
The sources were detected, and their properties 
were measured using tools provided in the Photutils 
package \citep{photutils}; the steps are 
described below.

Proper estimation of the background and associated 
uncertainty is vital for the detection 
of sources as well as the accurate estimation of 
source fluxes.
For each filter, a low-resolution background 
map was created by estimating the background 
in each mesh of a grid that covers the 
image using sigma-clipped statistics.
The mesh size should not be too small 
so that the background gets overestimated 
and source fluxes underestimated.
At the same time, the size should not be too 
large to miss the diffuse variation in the 
background. 
We chose 400$\times$400 pixels ($\sim$3$\arcmin$ 
$\times$ $\sim$3$\arcmin$) as the mesh size.
Sigma clipping is performed, up to 10 iterations, on each mesh by 
rejecting values outside 3 times the standard deviation from 
the median.
The background in each mesh is calculated as 
(2.5 $\times$ median) $-$ (1.5 $\times$ mean). 
However, if (mean $-$ median) $\div$ std $>$ 0.3, 
the median value is used as background.
Each mesh's background root mean square (RMS) 
is also estimated.
Then final background map is calculated 
by interpolating the low-resolution 
background map to match the image cutout size.
2D background and background RMS noise were 
estimated in this manner.

The NSR contains both extended and point-like sources. 
Therefore, we have used the image segmentation method of Photutils. 
In this method, sources are detected based on the criteria 
that there should be a minimum number of connected pixels 
(along edges or corners) that are each greater than 
a specified threshold value with a shape similar to the 
user-provided kernel.
The threshold was set as 1.5 times the background RMS 
off the background level. 
To separately identify detected sources, image pixels 
have integer labels assigned such that pixels with the 
same values are part of one source; such a map is 
called a segmentation image.

The segmentation image may contain overlapping sources 
detected as single sources. 
The deblending function provided in the Photutils 
package was used to separate such sources.
Photutils deblends sources in the segmentation image using a
combination of multi-thresholding and watershed segmentation. 
To deblend each source, 32 exponentially spaced thresholding 
levels were chosen between the minimum and maximum values of 
the source segment with the criteria that a deblended source 
should contain at least 0.1\% of the flux of the blended 
source. 

Compared to NUV, very few foreground stars are bright in 
the FUV channel.
Therefore, we used the F148W filter image to create a 
deblended source segmentation image that was also 
applied to NUV filter images. 
To smooth the noise and maximise the detection of sources 
with a shape similar to the telescope point spread 
function (PSF), the F148W image was filtered with a 
Gaussian kernel of FWHM of 1.4 arcseconds.
We removed sources with fluxes at or below the 
background level in the NUV filter images after applying 
the F148W segmentation image.
Finally, the centroid estimation and isophotal 
photometry were carried out for all the \totaldetections{} 
detected sources in each filter, where photometric 
fluxes were measured in counts per second (CPS). 
The CPS values are converted to flux in 
erg cm$^{-2}$ s$^{-1}$ \AA$^{-1}$ 
using the UVIT calibration information from 
\cite{tandon2017orbit} and 
\cite{tandon2020additional}.

The measured fluxes are affected by dust 
extinction in our galaxy;
the median colour excess along the line of sight 
to the NSR is 0.104 with a standard 
deviation of 0.006; we found no large-scale 
variations in the \cite{schlegel1998maps} 
colour excess map\footnote{
\cite{schlegel1998maps} map was obtained from 
\url{https://irsa.ipac.caltech.edu/applications/DUST}.
upon visual inspection}. 
\cite{schlafly2011measuring} suggests a 14\%
recalibration of \cite{schlegel1998maps} maps.
Therefore we took the colour excess value as 
0.089.
We then estimated
the flux extinction factors for each UVIT filter
with $R_v = 3.1$ following \cite{fitzpatrick1999correcting}, 
and the fluxes were corrected for the 
Galactic extinction.

Among the \totaldetections{} sources, there 
are foreground (stars) and background 
(distant galaxies) sources. 
Using Gaia distances from \cite{bailer2021estimating}, 
we identified \totalstars{} galactic sources, 
and they were removed from our list of sources  
(see Fig.~\ref{fig:sources}). 
The filtered catalogue was then compared against 
the SIMBAD astronomical database to remove 
those objects associated with a redshift 
above 0.003 (with 0.001826 being 
the Cen A redshift).
\farawayobjects{} sources were identified as 
distant galaxies and removed in this step. 
Our final catalogue thus consists of 
\sourcesremaining{} sources.
Since we do not have redshift information for 
the final NSR catalogue of \sourcesremaining{} sources
to conclusively ascertain their membership in
the northern star-forming region of Cen A,
our list of sources should be considered as 
candidates (while contamination has been 
controlled, there can still be intruders 
in this list).
The NSR source catalogue is publicly available
online\footnote{The catalogue can be accessed at \url{https://github.com/prajwel/Centaurus-A_NSR/blob/main/cena_nsr.fit}.}.
A sample table is shown in Table~\ref{tab:nsr_catalogue}.

HST observation of the Inner filament shows 
that both shock heated gas and young stellar 
populations are present along the structure 
\citep{crockett2012triggered}.
UVIT observation of the Inner filament also 
shows diffuse emission from the shock heated gas 
and concentrated emission from young stellar 
populations as point-like sources 
(see Fig.~\ref{fig:collage}, thumbnail 3).
Similar differences in emission features are 
observed in UV sources found near the Outer 
filament, with some being diffuse and others 
point-like.
Therefore, we assume that the diffuse UV emission 
comes from hot gas and point-like emission 
from Young stellar populations in the NSR. 
To differentiate the two, point-like sources in the NSR
source catalogue were identified using the DAOFIND
algorithm \citep{stetson1987daophot}.
We categorised point-like sources as sites of  
star formation and the remaining as diffuse 
emission sources.

\begin{table*}
\centering
\caption{A sample table from the NSR source catalogue 
         and their properties is shown. 
         RA and DEC are in degrees, the isophotal areas of NSR
         sources are in pc$^2$, and flux values are in 
         erg cm$^{-2}$ s$^{-1}$ \AA$^{-1}$.
         The whole catalogue can be accessed as a FITS table at
         \url{https://github.com/prajwel/Centaurus-A_NSR/blob/main/cena_nsr.fit}}.
\label{tab:nsr_catalogue}
\resizebox{\textwidth}{!}{%
\begin{tabular}{@{}rrrrrrrrrrr@{}}
\toprule
\multicolumn{1}{c}{RA\_J2000} &
  \multicolumn{1}{c}{DEC\_J2000} &
  \multicolumn{1}{c}{\begin{tabular}[c]{@{}c@{}}Area\end{tabular}} &
  \multicolumn{1}{c}{F148W flux} &
  \multicolumn{1}{c}{\begin{tabular}[c]{@{}c@{}}F148W flux\\ error\end{tabular}} &
  \multicolumn{1}{c}{N219M flux} &
  \multicolumn{1}{c}{\begin{tabular}[c]{@{}c@{}}N219M flux\\ error\end{tabular}} &
  \multicolumn{1}{c}{N245M flux} &
  \multicolumn{1}{c}{\begin{tabular}[c]{@{}c@{}}N245M flux\\ error\end{tabular}} &
  \multicolumn{1}{c}{N279N flux} &
  \multicolumn{1}{c}{\begin{tabular}[c]{@{}c@{}}N279N flux\\ error\end{tabular}} \\ \midrule
201.4154820 & -42.9330740 & 4.3644 & 6.8203E-17 & 5.1634E-18 & 8.3704E-17 & 9.8141E-18 & 6.1774E-17 & 6.1091E-18 & 6.0927E-17 & 7.6091E-18 \\
201.4189675 & -42.8136009 & 3.5977 & 5.1847E-17 & 4.5022E-18 & 6.1014E-17 & 8.4147E-18 & 5.3870E-17 & 5.2520E-18 & 4.3801E-17 & 6.1293E-18 \\
201.4198720 & -42.9460643 & 0.2949 & 1.5659E-18 & 1.0404E-18 & 2.5296E-18 & 2.1278E-18 & 3.6963E-18 & 1.3404E-18 & 5.0530E-18 & 2.0750E-18 \\
201.4203589 & -42.8501478 & 0.2949 & 3.3262E-18 & 1.2006E-18 & 8.6239E-19 & 1.4900E-18 & 1.6674E-18 & 8.2067E-19 & 1.6105E-18 & 1.3192E-18 \\
201.4270628 & -42.8393758 & 0.5308 & 5.9493E-18 & 1.6086E-18 & 2.1877E-18 & 1.8868E-18 & 1.6640E-18 & 8.5589E-19 & 6.3417E-18 & 2.4028E-18 \\
201.4308015 & -42.9838896 & 0.4718 & 5.4323E-18 & 1.5720E-18 & 1.3038E-20 & 1.2537E-18 & 1.8899E-18 & 1.3372E-18 & 2.0131E-18 & 1.6313E-18 \\
...         & ...         & ...    & ...        & ...        & ...        & ...        & ...        & ...        & ...        & ...        \\
201.7123104 & -42.7777452 & 2.1822 & 2.5372E-17 & 3.3228E-18 & 1.7310E-17 & 4.7520E-18 & 2.2578E-17 & 3.4386E-18 & 3.1436E-17 & 5.7720E-18 \\
201.7133760 & -42.7623472 & 0.9437 & 1.0718E-17 & 2.1710E-18 & 8.8000E-18 & 3.6816E-18 & 7.1867E-18 & 1.8518E-18 & 2.7222E-18 & 2.1107E-18 \\
201.7136741 & -42.9743428 & 1.8873 & 2.1280E-17 & 3.0789E-18 & 5.9550E-18 & 3.6667E-18 & 1.3223E-17 & 2.5992E-18 & 1.0956E-17 & 3.3486E-18 \\
201.7139861 & -42.7669313 & 0.4128 & 3.5732E-18 & 1.3393E-18 & 1.8493E-18 & 2.0008E-18 & 3.0481E-18 & 1.2401E-18 & 1.4284E-18 & 1.3460E-18 \\
201.7220010 & -42.7695121 & 0.3539 & 4.8505E-18 & 1.4020E-18 & 3.4040E-18 & 2.1360E-18 & 1.0263E-18 & 9.2078E-19 & 6.4431E-19 & 9.5644E-19 \\
201.7259025 & -42.7761188 & 4.0695 & 4.6224E-17 & 4.5108E-18 & 2.3046E-17 & 5.5320E-18 & 2.1584E-17 & 3.4222E-18 & 1.8171E-17 & 4.1800E-18 \\ \bottomrule
\end{tabular}%
}
\end{table*}

\section{Regions of NSR}

We divided the NSR further into Outer and
Inner regions.
The characteristics of the two regions are discussed
in the following sections.

\begin{figure}
	\includegraphics[width=\columnwidth]{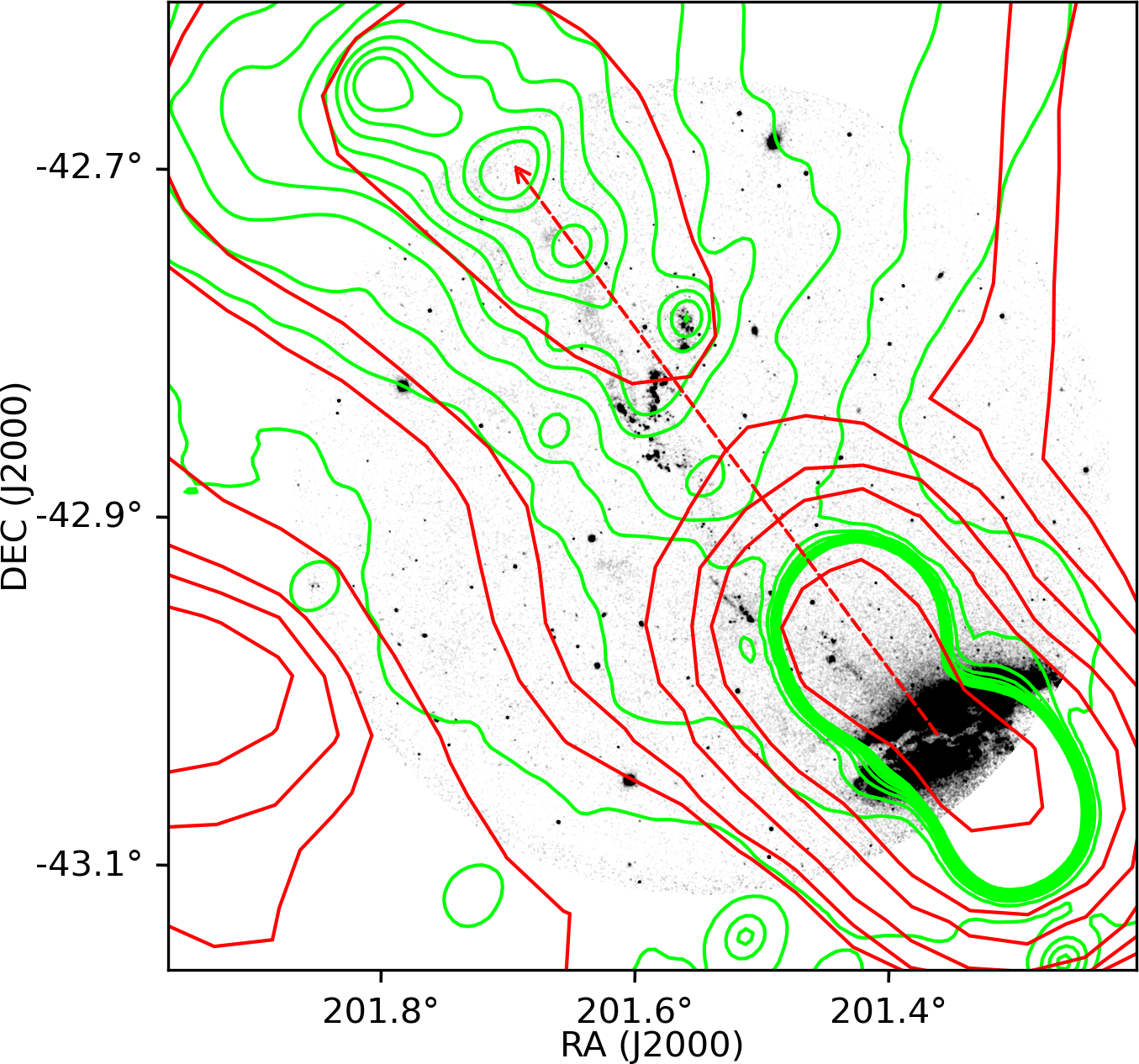}
    \caption{6.3 cm radio contours of \protect\cite{junkes1993radio} 
             with contour levels 
             0.1, 0.15, 0.3, 0.75, 1, 2, 5, 10, 40 Jy/beam
             and a half-power beamwidth of 4.3 arcminutes
             overlaid on the FUV F148W image. 
             The image has been processed to make the
             faint and diffuse features visible. 
             The red dashed line corresponds to a vector 
             connecting the AGN nucleus 
             to the core of 6.3 cm total power emission 
             in the NML.
             Green 185 MHz radio contours with contour 
             levels 0.2, 0.4, 0.6, 0.8, 1, 1.2, 1.4, 1.6, 1.8, 2 
             Jy/beam and a Gaussian restoring beam of width 
             1.5 $\times$ 1.2 arcminutes are from 
             \protect\cite{mckinley2022multi}.
             The broad emission observed by 
             \protect\cite{junkes1993radio} is likely from
             the radio knots in the NML being 
             smoothed due to a lack of sensitivity to 
             small-scale structures.}
    \label{fig:radio_emission}
\end{figure}

\subsection{Outer region}
\label{sec:outer_region}

To study the NSR sources observed in and around the Outer filament, 
we define a circular region of 4.5 arcminutes radius at 
RA = 201$\degr$.5884734, Dec = -42$\degr$.8144597 
(dashed circle in Fig.~\ref{fig:outer_region}).
The radius was chosen to encompass most
of the NSR sources near the Outer filament.
The double structure consisting of a V-shaped segment
and smaller elongated segment can be observed
here.
Part of the ribbon-like diffuse FUV emission 
noticed by \cite{neff2015complex} is also 
present.
\cite{mould2000jetCenA} suggested that the jet 
interacted with the gas cloud to trigger star 
formation.
A portion of the northern middle lobe (NML) is found 
in the Outer region.
The NML is part of Cen A radio structure 
classification in the literature, 
covering a section northward of Cen A up to 
$\sim$30 arcminutes \citep{israel1998centaurus}.
In the box area that we have selected for studying 
NSR (see Fig.~\ref{fig:showcase}), 
there exists NML and a smaller scale radio lobe
closer to the galaxy called the northern inner lobe
(NIL). 
While NML is present in the Outer region, 
NIL is found in the Inner region (see next section).

There has been no detection of a southern middle lobe 
or SML \citep{neff2015radio}.
The observations by \cite{morganti1999centaurus} revealed
a large-scale jet that connects the NML and NIL.
They proposed that NML could be the result of
a precessing jet that underwent an interaction
with the environment on the northern side of Cen A. 
\cite{saxton2001centaurus} considered the scenario
of NML being a buoyant bubble of plasma generated
by an intermittently active jet. 
According to \cite{kraft2009jet}, NML was initially
formed from a previous nuclear outburst and is 
being re-energised by the current outburst. 
\cite{2010gopal} described how the radio jet 
could have interacted with the gas cloud to produce
NML.
To address the various short-lived structures seen 
in and around the NML, \cite{neff2015complex} proposed
that an AGN-augmented wind could be present.
Contrary to what is observed and explained by
\cite{morganti1999centaurus}
as a currently active 'large-scale jet', 
radio observations by \cite{neff2015radio} do not
detect any structure connecting NIL and NML.  
\cite{mckinley2018jet} detected a 
collimated structure at the site of 'large-scale jet'.
The latest observations by \cite{mckinley2022multi} 
disfavours the existence of a presently active 
'large-scale jet'.

Regardless of whether a collimated jet 
now exists connecting NIL to NML and 
AGN-augmented wind currently energises the 
observed structures around Cen A,
we consider it likely that a jet activity could 
have been present extending to NML in the past 
for the following reasons:
(a) The presence of a double structure in the Outer
region---the observed UV sources show a gap in 
its distribution, and it splits the structure into 
a V-shaped segment and a smaller elongated segment. 
Such double structures are observed in association
with other cases of jet induced star formation
(See Table~\ref{tab:jetinduced}).
(b) The spatially localised anomalous gas kinematics
in a large gas cloud found near the Outer filament 
is explained as arising due to jet interaction 
\citep{oosterloo2005anomalous}.
(c) The wind depends on an ongoing starburst in the
galaxy, but the giant radio lobes seen on the north and 
south sides of Cen A have an age of the order of 
1 Gyr \citep{eilek2014dynamicmodelRadio}. 
Simulations predict that it is less likely for a 
starburst to be older than 500 Myr \citep{di2008frequency}. 

Therefore, we adopt the interpretation of \cite{2010gopal},
which explains the origin of NML as arising due to 
jet interaction with the gas cloud. 
Observations by \cite{neff2015radio} and \cite{mckinley2022multi}
show that the NML has multiple radio knots.
X-ray knots were observed in the NML by \cite{kraft2009jet}.
Previous authors noted that the radio and X-ray 
knots may dissipate within a few Myr and requires 
a supply of energy to maintain them.
Based on energetic arguments, \cite{mckinley2022multi} 
suggested that the AGN jet is a necessary part
of explaining the observed features in the NML.
We assume that the radio and X-ray knots 
in the NML were created by a former jet when a 
gas cloud would have come into contact with it. 
We believe this is a reasonable assumption based on
existing observations of the NML.

To find the potential path of this past jet activity
through the Outer region,
we used the radio continuum data of \cite{junkes1993radio}
and estimated the centroid for the sigma clipped core 
profile of the radio emission in the NML 
(see Fig.~\ref{fig:radio_emission}, the bottom
part of the NML can be seen extending into the Outer
region).
Then, extending from the AGN nucleus, we created a 
vector that goes through the NML centroid. 
This vector has a position angle of 37.4 degrees. 
The \sourcesremaining{} NSR sources (\sfsources{} 
star-forming and \diffusesources{} diffuse sources) 
are shown in Fig.~\ref{fig:sources}, along with the vector.
The path along the vector appears to be relatively 
devoid of NSR sources.
The absence of NSR sources along the vector path is 
especially pronounced in the Outer region.
The vector goes through the gap splitting the observed 
double structure into two segments. 
We consider the vector as the potential path 
of the past jet activity through NML.
Our vector is aligned to the first two radio knots
observed by \cite{mckinley2022multi}.
We caution that the vector represents a 
hypothetical path; it may be difficult to ascertain
whether the jet travelled along this path.

The ALMA observations by \cite{salome2017inefficient} 
reveals that there is also a lack of molecular clouds 
along the vector path (see Fig.~\ref{fig:outer_region}). 
Detections from the ALMA CO map are overlaid in cyan colour. 
The radii of circles are proportional to the cloud masses.
Most molecular clouds are located along with the 
ribbon-like diffuse FUV emission region.
If the diffuse UV emission comes from hot gas, 
the observed cluster of molecular clouds could be far 
from the hot gas, though they appear close in projection.




\subsection{Inner region}

\begin{figure}
	\includegraphics[width=\columnwidth]{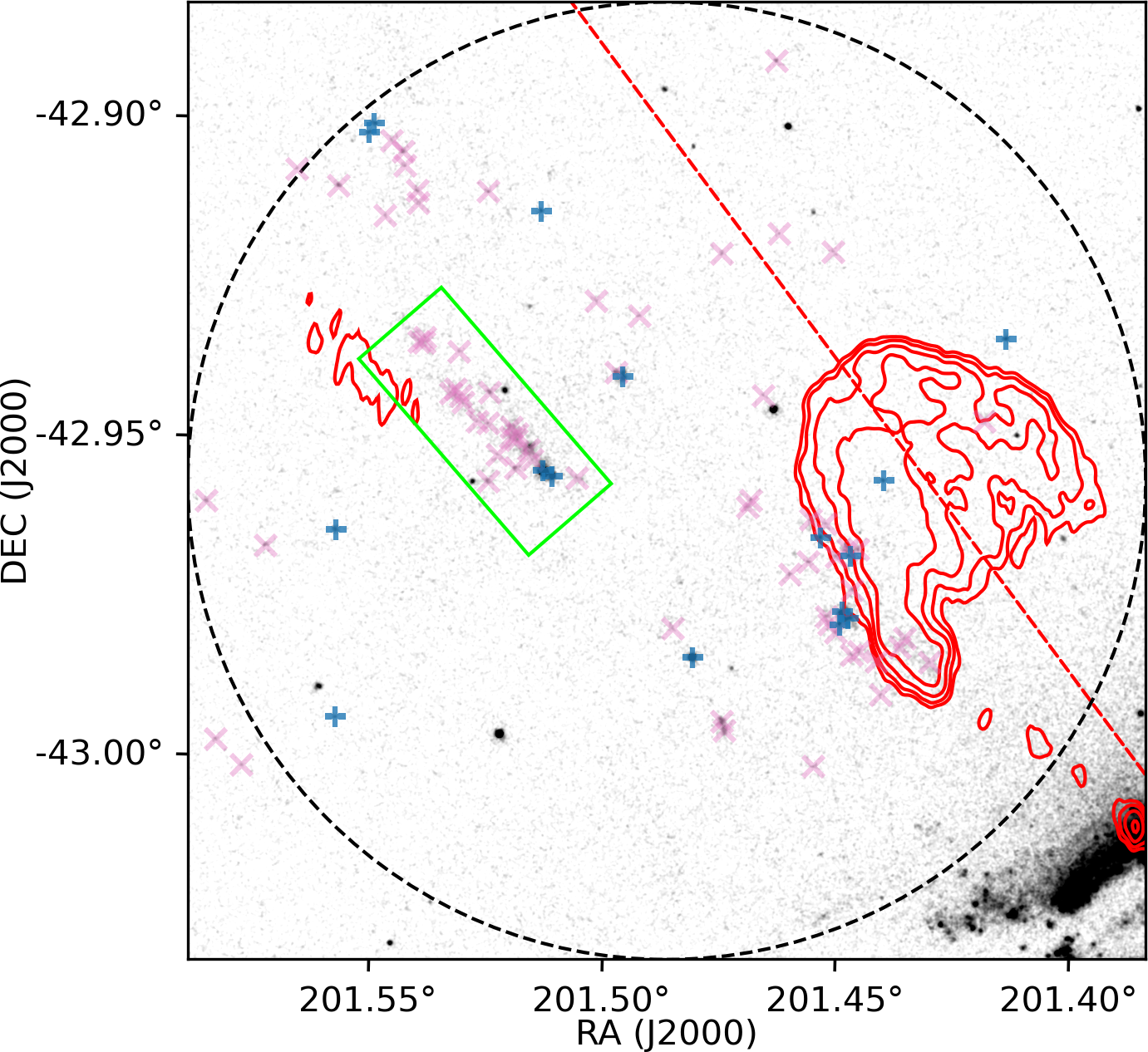}
    \caption{The greyscale F148W image shows 
             the Inner region.
             Blue crosses are star-forming sources, and 
             pink X-shaped markers depict diffuse sources.
             The red dashed line is the vector path.
             The green box shows the location of
             the Inner filament.
             The VLA 1.55 GHz radio contours are overlaid (red) 
             on the figure with contour
             levels 0.060, 0.097, 0.167, 0.287 Jy/beam 
             and a half-power beamwidth of 4.3 arcseconds. 
             The radio feature close to the Inner filament has only the lowest contour level; 
             therefore, this might be an artefact.
             VLA image credit: NRAO/VLA Archive Survey, (c) 2005-2009 AUI/NRAO.}
    \label{fig:inner_region}
\end{figure}

Compared to the Outer filament, the Inner filament 
has fewer NSR sources. 
For a comparative analysis, we defined an expanded 
region including the Inner filament with a circle 
of 4.5 arcminutes radius (same as the Outer region) at 
RA = 201$\degr$.4875502, Dec = -42$\degr$.9560958.
The inner filament sources can be seen as an
elongated structure.
Very young stars (1--4 Myr) have been observed in 
the Inner filament by \cite{crockett2012triggered}, 
and they suggested that a jet cocoon driven bow 
shock could have triggered star formation.
Close to the Inner filament, backflows from the jet have
been found \citep{hamer2015muse}. 
The FUV F148W image with VLA 1.55 GHz contours reveals 
the complex structure of the Inner region 
where present jet activity is observed
(see Fig.~\ref{fig:inner_region}). 
According to the Cen A radio structure classification,
the radio lobe seen in the Inner region is called the
northern inner lobe (NIL).

\begin{figure}
	\includegraphics[width=\columnwidth]{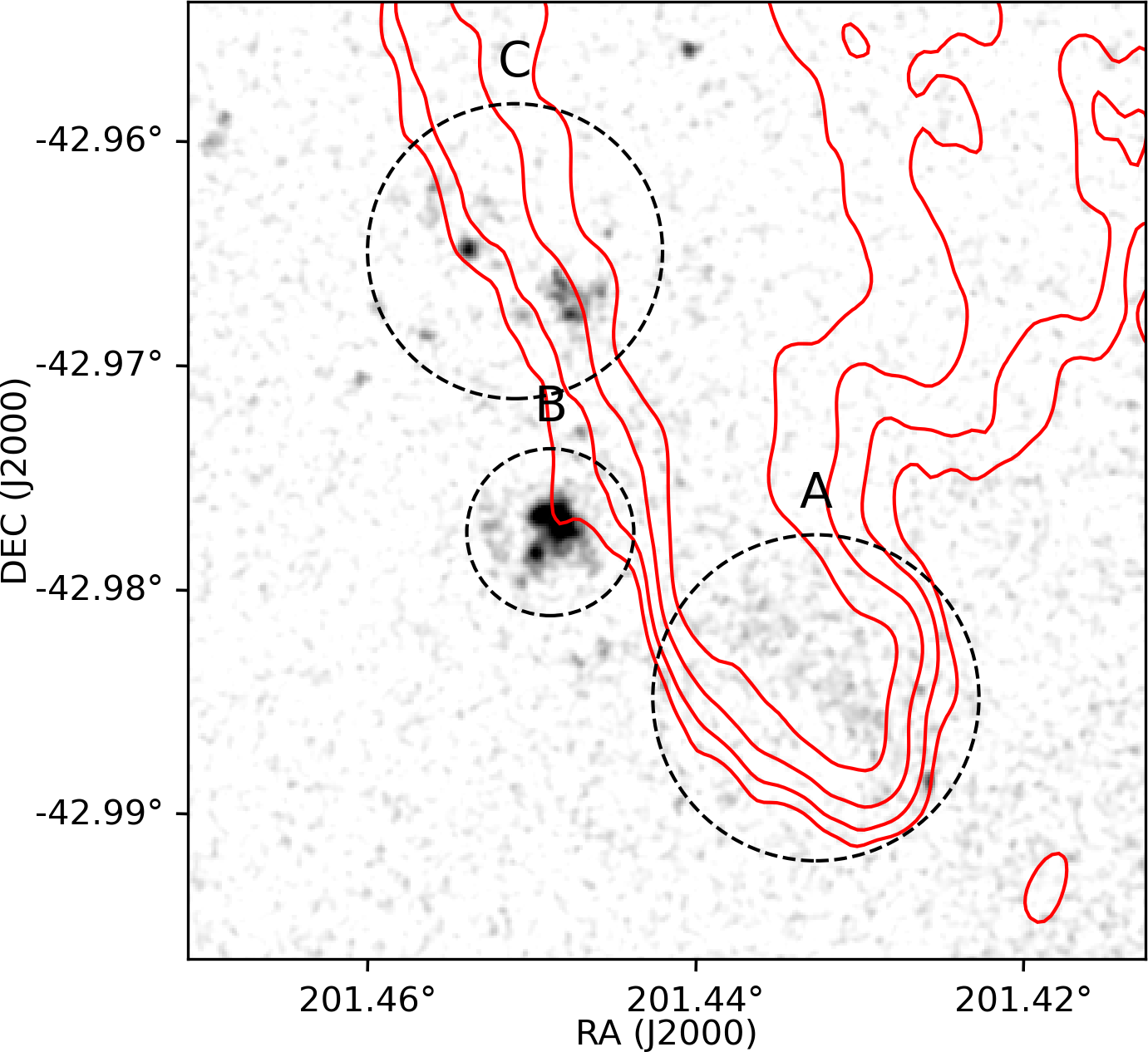}
    \caption{A section of Figure \ref{fig:inner_region}  
             containing the lower part of NIL is shown here.
             The image was convolved with a Gaussian kernel
             with FWHM comparable to the PSF. 
             Dashed circles B and C represent candidate 
             star-forming structures, and A denote diffuse emission.}
    \label{fig:new_regions}
\end{figure}

\begin{figure}
	\includegraphics[width=\columnwidth]{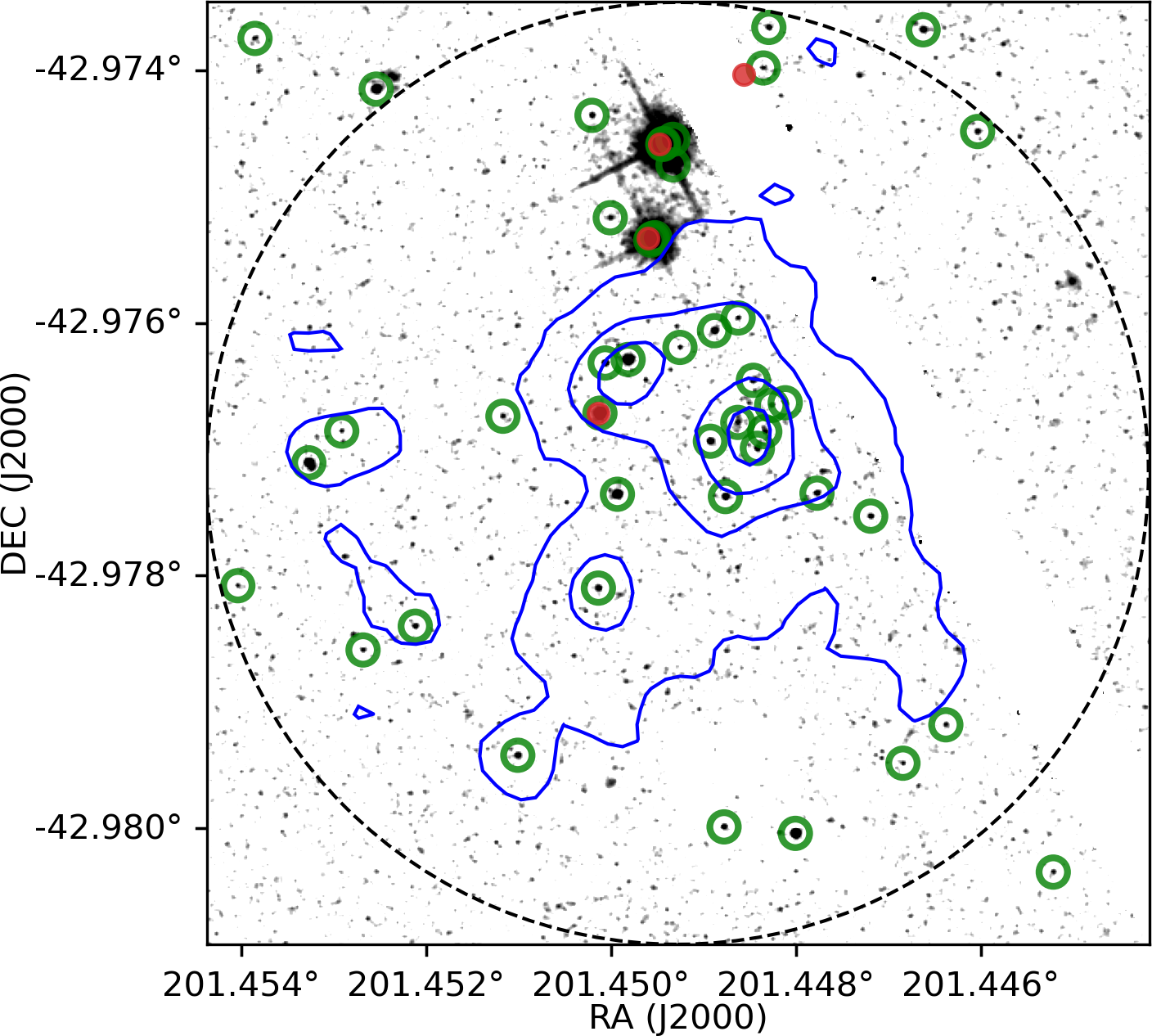}
    \caption{HST F606W image of the candidate site B 
             is shown.
            The white diagonal stripe on the right
             side of the figure represents the 
             $\sim$50 pixel gap in the ACS
             detector.
             The dashed circle occupies the 
             same sky region as the dashed circle B in
             Fig.~\ref{fig:new_regions}. 
             The UVIT F148W contours are overlaid (blue) 
             with contour levels 0.00012, 0.00028, 0.00044, 
             0.0006 CPS.
             The green circles mark the point-like
             sources detected using DAOFIND.
             The red-filled circles show the Gaia detected foreground 
             stars \citep{bailer2021estimating}.}
    \label{fig:subregion_B}
\end{figure}

\begin{figure}
	\includegraphics[width=\columnwidth]{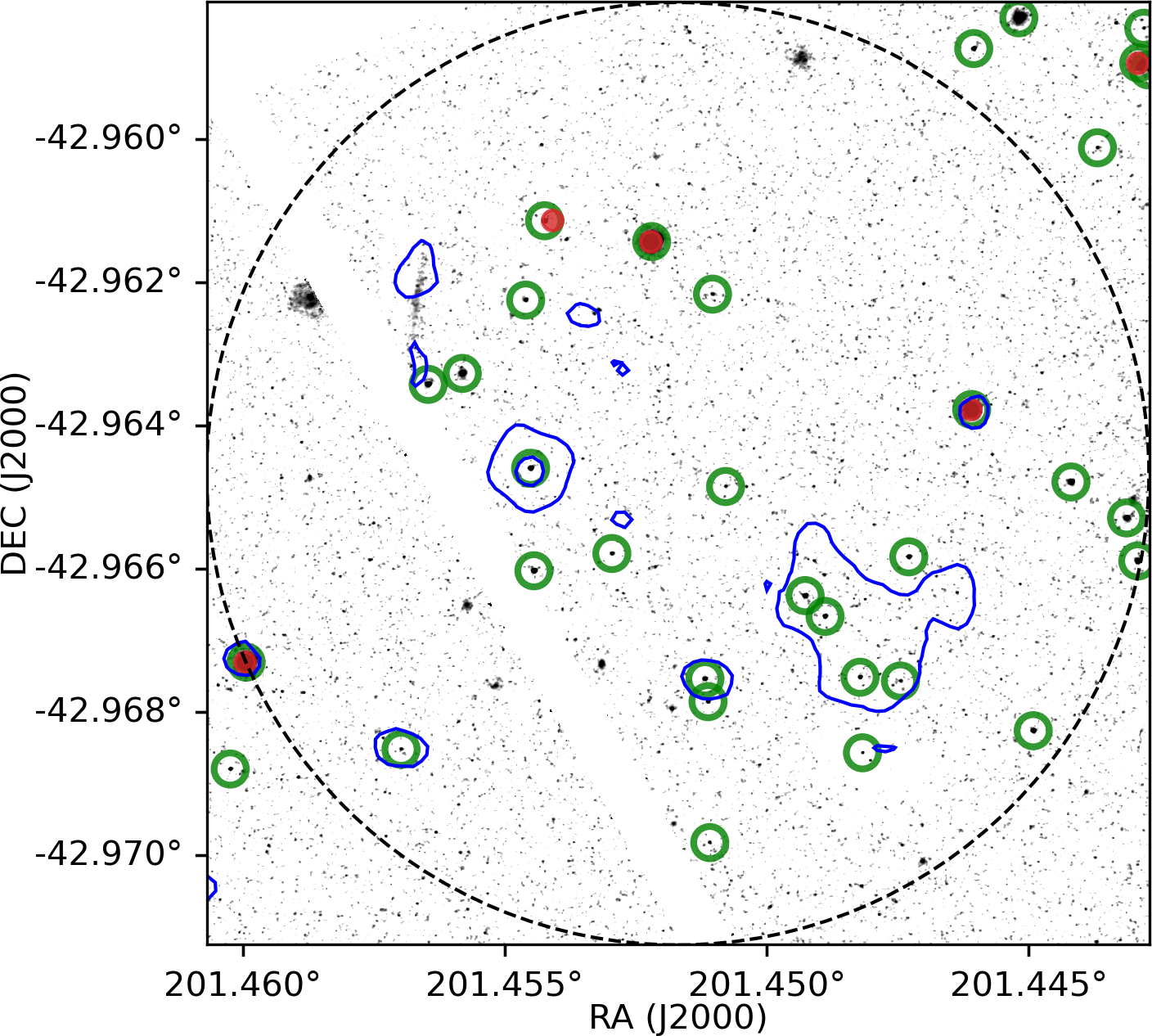}
        \caption{HST F606W image of the candidate site C 
             is shown. 
             The white diagonal stripe on the left 
             represents the 
             $\sim$50 pixel gap in the ACS
             detector.
             Additionally, the ACS detector did not cover 
             a small section near the top left corner. 
             This section is also shown in white colour.
             The dashed circle occupies the 
             same sky region as the dashed circle C in
             Fig.~\ref{fig:new_regions}. 
             The UVIT F148W contours are overlaid (blue) 
             with contour levels 0.00012 and 0.00028 CPS.
             The green circles mark the point-like
             sources detected using DAOFIND.
             The red-filled circles show the Gaia detected foreground 
             stars \citep{bailer2021estimating}.}
    \label{fig:subregion_C}
\end{figure}

We have identified candidate sites of star formation 
that could be related to AGN activity
(dashed circles B and C in Fig.~\ref{fig:new_regions}).
They appear where the collimated radio beam in
the NIL starts to diverge. 
The candidate site B is mentioned in the literature
as DF 6 and has been noted for its alignment with the jet 
\citep{dufour1978inner, graham2002star}.
We found Advanced Camera for Surveys (ACS) 
F606W observations of the candidate sites from the HST
archive (HST programme 10597, PI: A. Jordan).
F606W is a broad V-band filter that includes the
H-alpha emission line. 
The F606W images of the candidate sites are shown
in Figs \ref{fig:subregion_B} and \ref{fig:subregion_C}.
Point-like sources are seen in the F606W image 
at locations where FUV emission is observed.
The green circles in Figs \ref{fig:subregion_B} 
and \ref{fig:subregion_C} mark the point-like
sources detected using DAOFIND.
In Fig.~\ref{fig:subregion_B}, a cluster of point-like
sources is seen aligned with the core of the FUV
emission contours.
The alignment of candidate sites B and C with NIL 
makes a case for jet induced star formation at 
these locations.

\begin{figure}
	\includegraphics[width=\columnwidth]{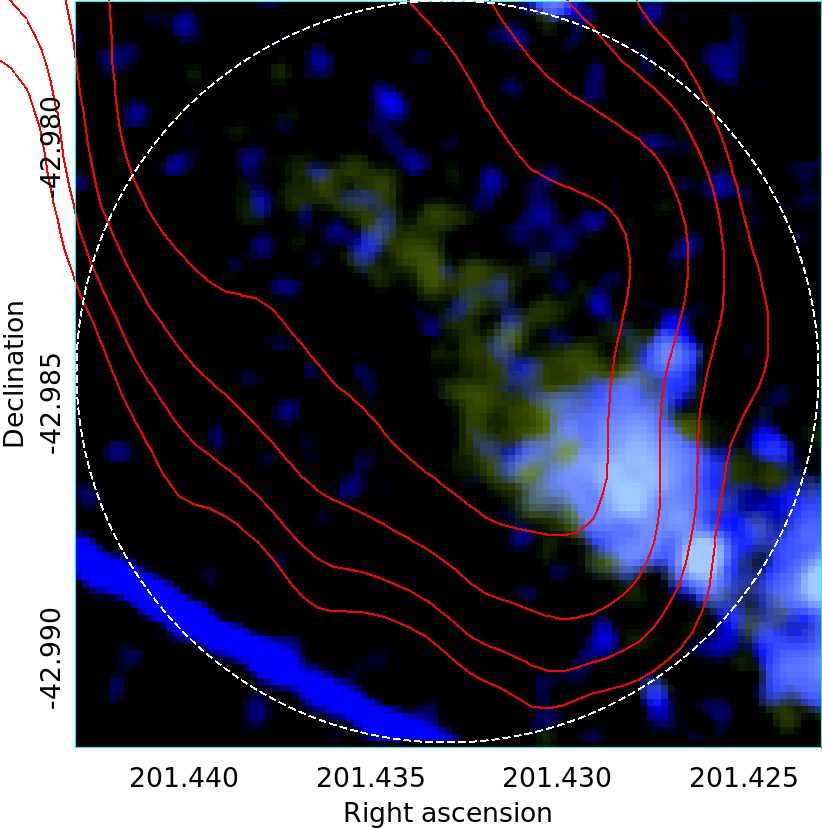}
        \caption{\textit{Chandra} ACIS-I image of 
                 candidate site A is shown with soft band flux
                 (0.5-1.2 keV) in yellow and hard band flux
                 (2.0-7.0 keV) in blue. 
                 The diagonal blue stripe near the
                 bottom left corner is due to the read-out
                 streak in the hard band 
                 \citep{mccollough2005impact}.
                 The dashed circle occupies the same sky 
                 region as the dashed circle A in
                 Fig.~\ref{fig:new_regions}.
                 The VLA 1.55 GHz radio contours are overlaid 
                 (red) with contour levels 0.060, 0.097, 0.167, 
                 0.287 Jy/beam.}
    \label{fig:subregion_A}
\end{figure}

There are also faint and diffuse UV sources 
padded along the radio beam path (dashed 
circle A in Fig.~\ref{fig:new_regions}).
Using the SciServer system and CIAO 4.13 (CALDB 4.9.6), 
we merged all available \textit{Chandra} ACIS-I 
observations of Cen A \citep{taghizadeh2020sciserver, 
fruscione2006ciao}.
\textit{Chandra} ACIS-I image of candidate site A is shown 
in Fig.~\ref{fig:subregion_A} with soft band flux (0.5-1.2 keV) 
in yellow and hard band flux (2.0-7.0 keV) in blue colours. 
If we compare Figs \ref{fig:new_regions} and \ref{fig:subregion_A},
hard band flux is missing at locations where diffuse 
UV emission is observed. 
Also, we did not observe any point-like sources in 
HST F606W observations within site A.
Noting these aspects, we propose that diffuse 
emission could be coming from the hot gas 
present at this location.

\subsection{Stellar population synthesis}
\label{sec:age}
We used the evolutionary synthesis code
Starburst99 v7.0.1 to analyse the NSR sources 
with recent star formation in both regions
\citep{leitherer1999starburst99, 
       vazquez2005optimization, 
       leitherer2010library, 
       leitherer2014effects}. 
Starburst99 has already been used in studies 
with UVIT data \citep{mondal2018uvit}. 
\cite{salome2016star} noted that Outer 
and Inner filaments have high oxygen abundances 
from the Multi-Unit Spectroscopic Explorer 
(MUSE) data analysis. 
While the MUSE observations cover only 
small sections of the Outer and Inner regions 
(2 arcminute square section on the V-shaped segment
of the Outer region and 1 arcminute square covering the
inner filament), \cite{salome2016star} found 
no major metallicity differences between the filaments
and no major metallicity gradients along each of the 
filaments.
Therefore, we have assumed solar metallicity in both
regions.
We used a simulated simple stellar population that had 
its total stellar mass distributed in a Kroupa IMF with 
instantaneous star formation evolved through Geneva 
isochrone tracks (solar metallicity; Z = 0.014) 
to compare our observations.
To obtain bounds on the effects of massive star rotation,
we used the 0\% and 40\% of the break-up velocity isochrone
models provided in Starburst99.
Following the notation in \cite{leitherer2014effects},
we denote the 0\% rotation model with solar metallicity as 
v00-h and the 40\% model as v40-h.  

\begin{figure}
	\includegraphics[width=\columnwidth]{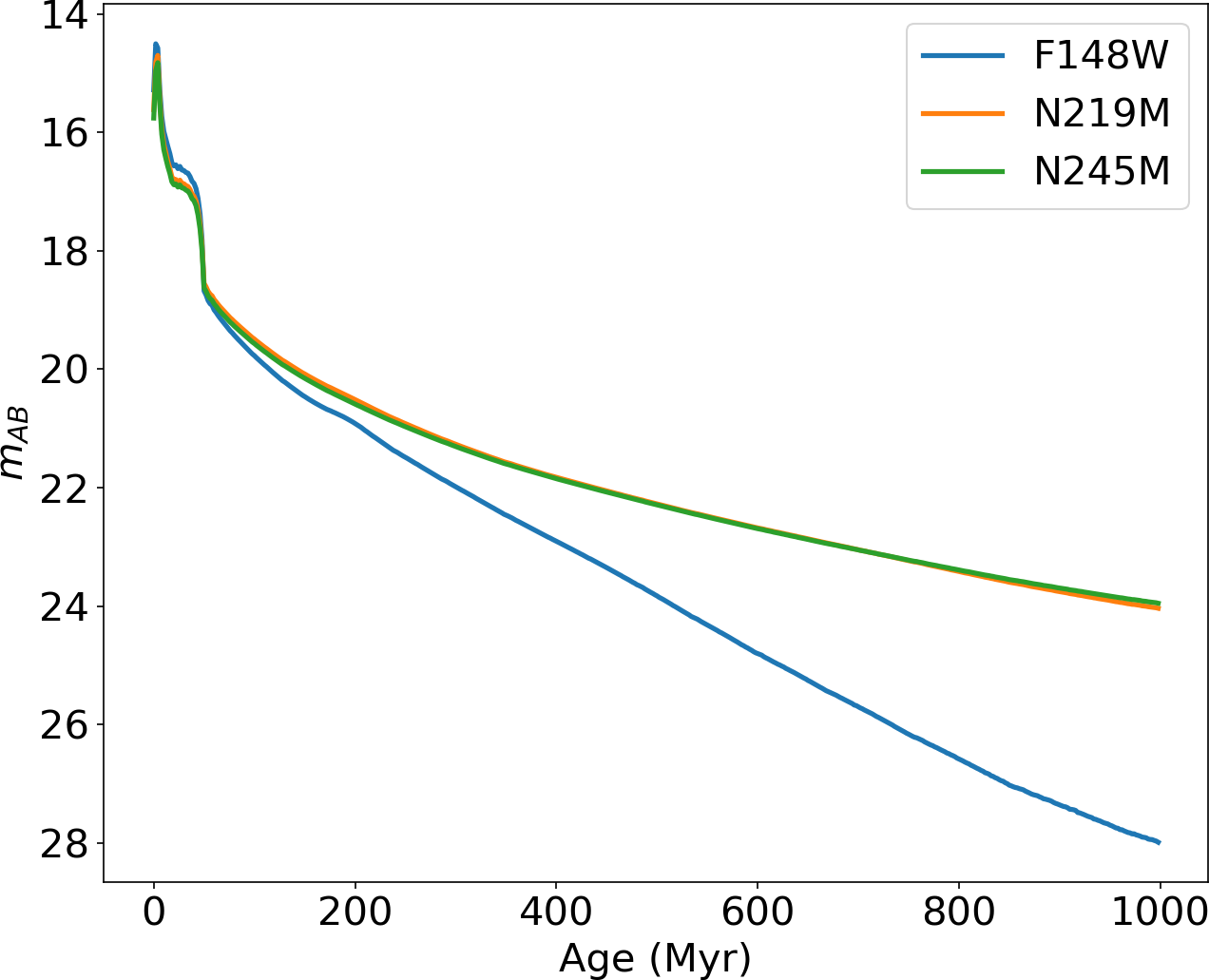}
    \caption{The plot shows the UVIT filter AB magnitudes
             simulated using Starburst99.
             The F148W, N219M, and N245M magnitudes are 
             shown for the 0-1 Gyr range. 
             The simulation was generated using the 
             v00-h model and
             {\boldmath $10^5 M_{\odot}$} total stellar mass.}
    \label{fig:simulated_colors}
\end{figure}
\begin{figure}
	\includegraphics[width=\columnwidth]{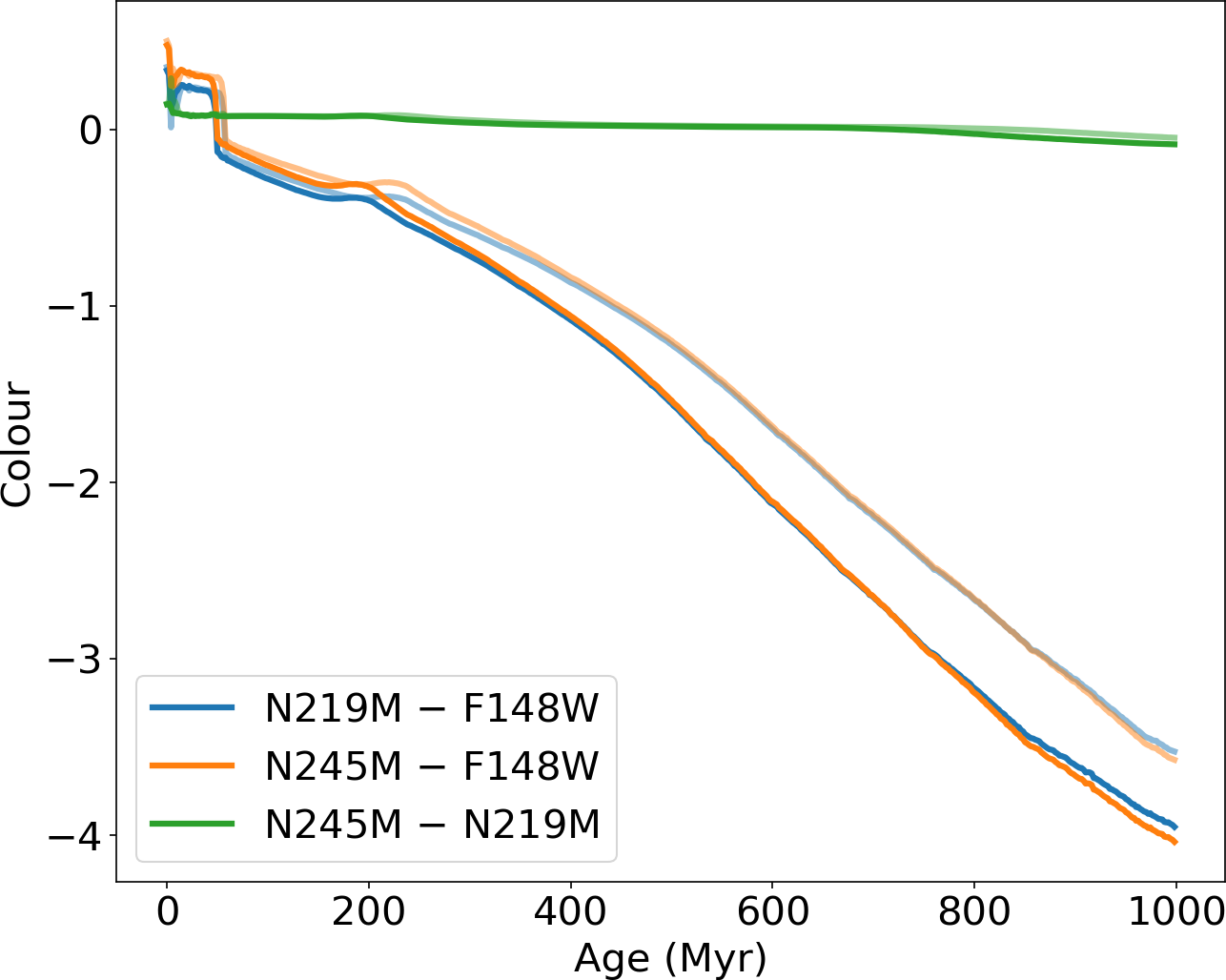}
    \caption{The plot shows the UVIT filter  
             colours simulated using Starburst99.
             The N219M $-$ F148W (blue), 
             N245M $-$ F148W (orange), 
             and N245M $-$ N219M (green) colours have been 
             estimated for the 0-1 Gyr range. 
             Strong and faint lines correspond to 
             simulations with v00-h and v40-h, respectively.}
    \label{fig:age_attenuation}
\end{figure}

The method followed for the analysis is explained below.
Using the simulated high-resolution UV spectra 
spanning ages up to 1 Gyr,
we generated a table of UV luminosities 
in UVIT filters by convolving simulated spectra 
with filter effective area functions. 
The simulated luminosity $L(\lambda_{\text{mean}}$) 
in erg s$^{-1}$ \AA$^{-1}$
was calculated as follows:

\begin{equation}
L(\lambda_{\text{mean}}) = 
\frac{\int L(\lambda)\ EA(\lambda)\ d\lambda}
     {\int EA(\lambda)\ d\lambda}
\end{equation}

where $L(\lambda)$ is the starburst99 spectra and 
$EA(\lambda)$ is the effective area function given in
Table 4 of \cite{tandon2020additional}.
The $\lambda_{\text{mean}}$ is the mean of
effective area weighted wavelengths (defined in equation
3 of \citealt{tandon2017orbit}).
The luminosities were then 
converted to AB magnitudes.
The simulated magnitudes are shown in 
Fig.~\ref{fig:simulated_colors} for F148W, N219M, and
N245M bands with the v00-h model and 
$10^5 M_{\odot}$ total stellar mass.
The observed FUV $-$ NUV colour 
can be compared with the simulated colour 
to estimate the age of the UV emitting 
population of young stars.
From our list of \sfsources{} NSR star-forming 
sources, 61 sources lie inside the Outer region and 
16 are found in the Inner region.
Fig.~\ref{fig:age_attenuation} shows ratios of the simulated 
luminosity values between UVIT filters; output from simulations
using both v00-h and v40-h models are shown. 
However, dust attenuation could affect UV 
observations in the Outer and Inner regions. 
\cite{auld2012herschel} observed dust clouds around 
Cen A and interpreted that a merger with a 
late-type gas-rich galaxy could have taken place. 
\cite{santoro2016embedded} observed a small section 
of the Outer filament using MUSE 
and derived colour excess values 
from $\text{H}\alpha/\text{H}\beta$ 
ratio for two star-forming sources.
We found that the N245M - N219M colour
remain approximately constant (see Fig.~\ref{fig:age_attenuation}). 
After assuming an attenuation law with $R_v = 4.05$
from \cite{calzetti2000dust}, the simulated and observed 
colours were compared to estimate the 
colour excess (see appendix~\ref{sec:colour_excess}).

\begin{figure}
	\includegraphics[width=\columnwidth]{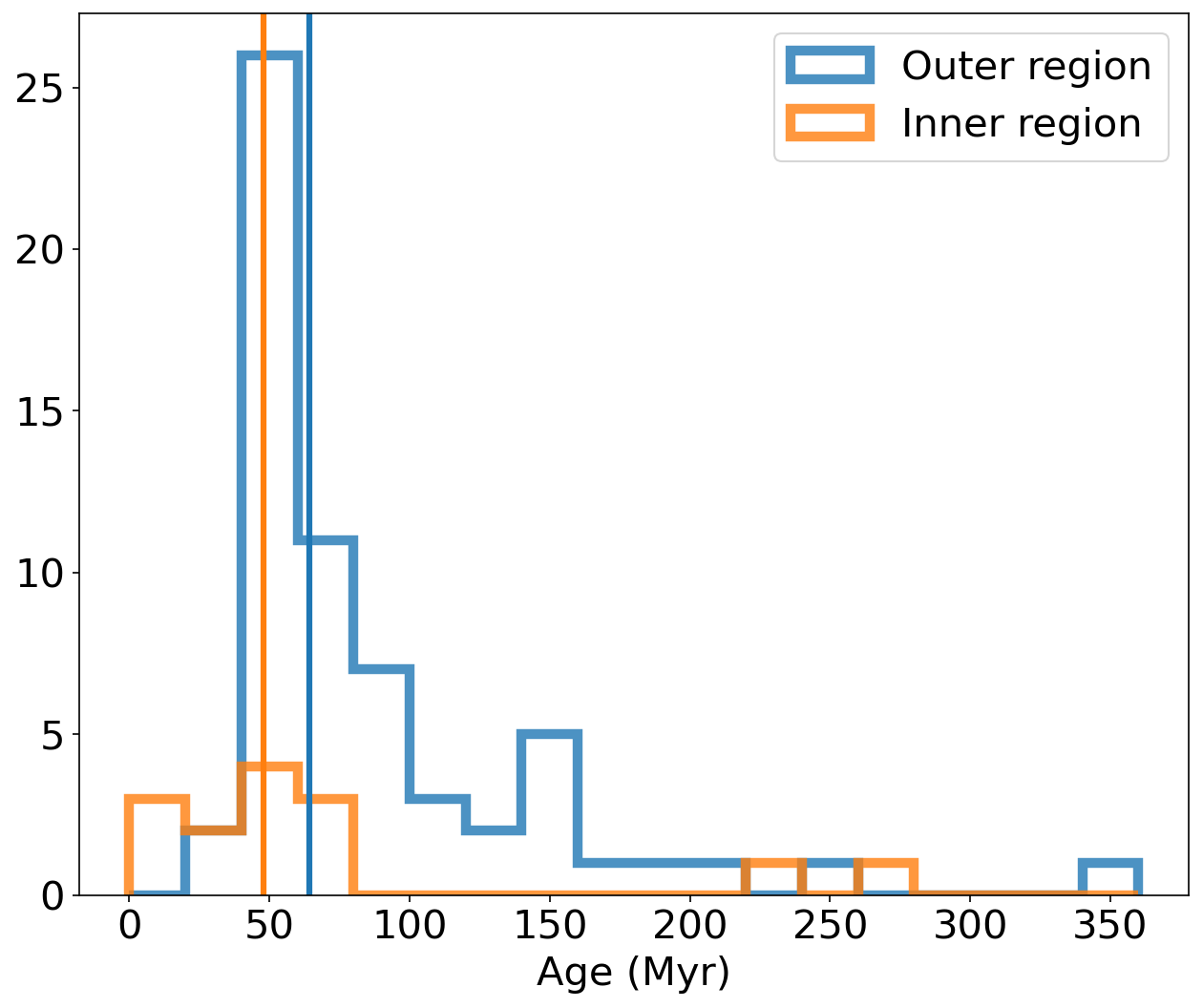}
    \caption{The age distribution of the Outer and Inner regions.
             Each histogram bin has a width of 20 Myr.
             The vertical blue and orange lines represent the 
             median values of the Outer and Inner region 
             distributions, respectively (see Table~\ref{tab:age_summary}).}
    \label{fig:age_distribution}
\end{figure}

The flux and magnitude values were 
corrected for dust attenuation. 
Then, to estimate the ages of the sources, 
the observed N219M $-$ F148W and 
N245M $-$ F148W colours
were compared with simulated colours 
from the v00-h model.
The mean of the two estimates from 
N219M $-$ F148W and N245M $-$ F148W colours
was taken as the final age value for each source.
The colours and estimated ages for the Outer 
and Inner regions are given in Tables~\ref{tab:inner_region_catalog}
\& \ref{tab:outer_region_catalog}, respectively.
See Table~\ref{tab:age_summary} for a 
summary of results in both regions.
N279N is a narrow bandwidth filter 
($\sim$7 times less sensitive than N245M); 
hence observations
in this filter were not used for age calculations. 
Fig.~\ref{fig:age_distribution} shows the age 
distribution of the Outer and Inner regions.
The V40-h model also gives comparable age 
estimates.
Compared to the Outer region, where most sources are part of 
the Outer filament, the Inner region contains a more sparse 
distribution of ages. 



\begin{table}
\centering
\caption{Summary of colour excess and age estimations. 
Median values  
of the age distributions are given.
The age spread (Myr) row gives the low and high limits
to the age distribution.}
\label{tab:age_summary}
\begin{tabular}{@{}lrlrl@{}}
\toprule
\multirow{2}{*}{} & \multicolumn{2}{c}{\multirow{2}{*}{Outer region}} & \multicolumn{2}{c}{\multirow{2}{*}{Inner region}} \\
                  & \multicolumn{2}{c}{}               & \multicolumn{2}{c}{}               \\ \midrule
E(B-V)            & \multicolumn{2}{r}{\outerexcess{}} & \multicolumn{2}{r}{\innerexcess{}} \\
Median age (Myr)  & \multicolumn{2}{r}{\outermedian{}} & \multicolumn{2}{r}{\innermedian{}} \\
Age spread (Myr)  & \multicolumn{2}{r}{\outerspread{}} & \multicolumn{2}{r}{\innerspread{}} \\ \bottomrule
\end{tabular}
\end{table}

\section{Discussion}

The AGN jet enabled star formation is emerging 
as a viable scenario in recent observations. 
The star formation associated with some of the 
filaments of NGC 1275 has been explained as arising 
due to AGN jet driven buoyant bubbles 
\citep{ngc1275canning2010star, ngc1275canning2014filamentary}. 
Citing the simulation of \cite{gaibler2012jet}, a bow shock 
created from the jet propagation has been invoked by 
\cite{ngc7252george2018uvit} to describe the observed 
star-forming ring of NGC 7252. 
For 3C 285, there are signatures of a recent merger in
the parent galaxy \citep{salome2015jet}.
In Minkowski's object (MO), a double structure 
that straddles the AGN jet has been observed 
\citep{croft2006minkowski}.
We have compiled a list of some of the known 
cases of jet induced star formation and their 
properties (see Table~\ref{tab:jetinduced}). 
Interestingly, almost all are associated 
with merger activity. 
Another aspect is the ubiquitous bifurcated profile;
the jet induced star-forming sources are loosely 
organised into a double structure.
In most cases, there is an apparent dichotomy between
the sizes of this double structure, with one being
larger than the other. 
This asymmetry in sizes appears to be a common 
feature among jet-cloud interactions.
We consider the possibility that when the jet first 
encounters the gas cloud, it is unlikely to 
pass exactly bisecting the gas cloud,which may lead
to the observed morphology.

\begin{figure}
	\includegraphics[width=\columnwidth]{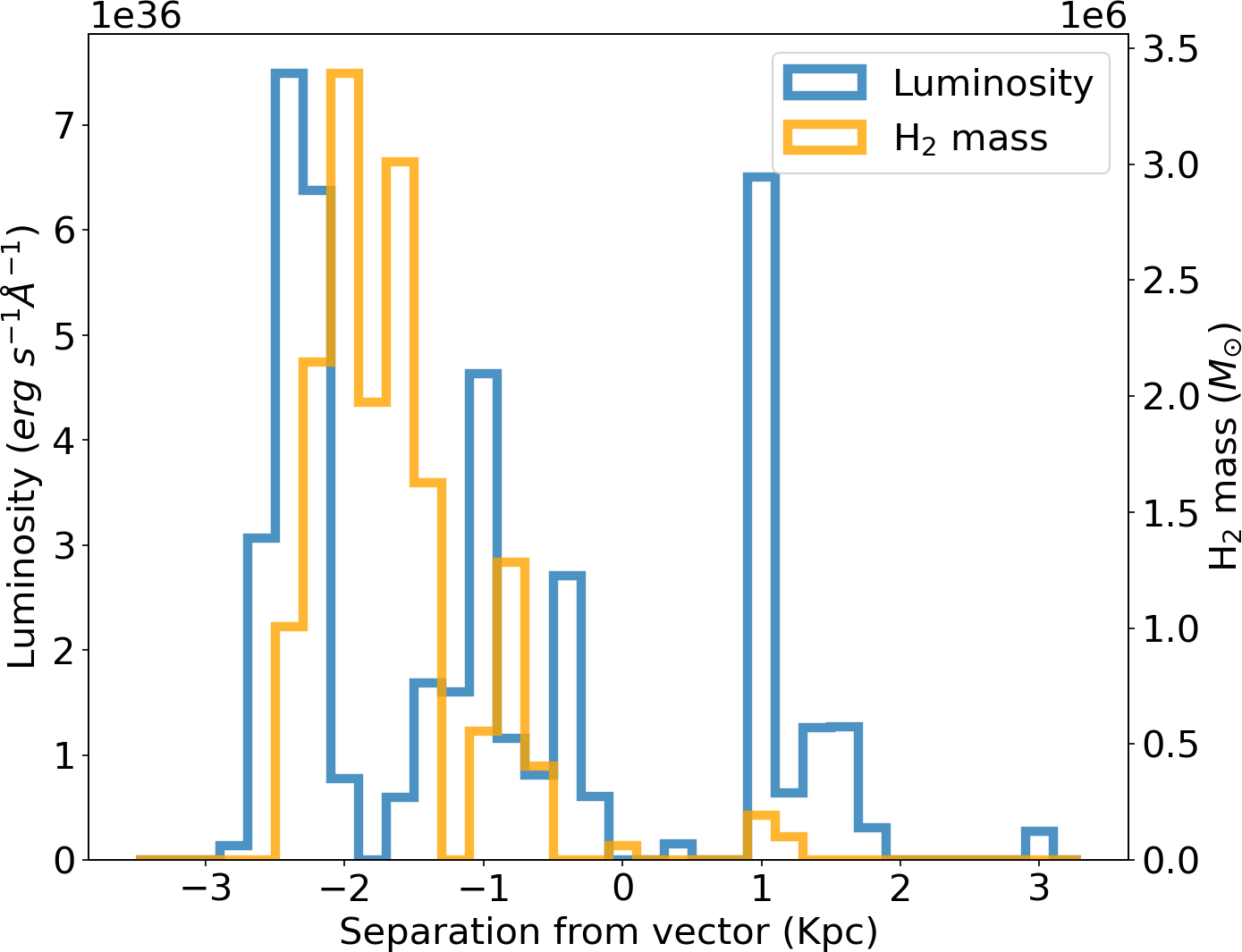}
    \caption{The vector separation in kpc for 
             star-forming sources and 
             molecular clouds part of the Outer region. 
             Each bin in the histogram represents a width of 0.2 kpc.
             The UV sources are represented using 
             the blue histogram, 
             and the left Y-axis represents the total F148W luminosity 
             in each bin for those sources.
             Similarly, molecular clouds in the 
             orange histogram show the cloud mass 
             in each bin on the right Y-axis.}
    \label{fig:separation}
\end{figure}

Similar to other sources, the jet-cloud interaction
could be responsible for the observed bifurcation 
in the Outer region of Cen A.
To further probe this feature in the Outer region, we 
estimated the angular separation from the vector for 
all star-forming sources and molecular clouds.
The angular separations were then converted to kpc, and
Fig.~\ref{fig:separation} shows the vector separation 
for sources and molecular clouds part of the Outer region;
each bin is of 0.2 kpc width with the height of the bar 
representing the total F148W luminosity (blue) and 
cloud mass (orange). 
Sources seem to be absent in a $\sim$1 kpc 
width northwest of the vector.
The observed gap could be the area through which 
the jet may have passed and triggered star formation 
on both sides.
Therefore, as previously pointed out by various 
studies, our observations of the Outer region 
favours a scenario where jet-cloud interaction 
led to triggered star formation.

Compared to the star-forming structure observed in
the Outer region, the Inner region has small 
structures.
Such small star-forming structures may also be 
present in other 
jet-induced star formation cases, but the
emphasis of studies has been chiefly concentrated
on the major structure (for example, see figure 5a 
in \citealt{van1993induced3c285}).
The star-forming sources in the Inner 
filament have been attributed to a bow shock 
\citep{crockett2012triggered}.
jet backflows are observed in the area 
\citep{hamer2015muse}.
If not an imaging artefact, the relic radio emission 
seen north-east in Fig.~\ref{fig:inner_region} 
could be associated with the jet activity that 
caused the observed features in the Inner filament. 
Apart from the known Inner filament sources, 
we have identified new sites of star formation 
(see B and C in Fig.~\ref{fig:new_regions}).
They are seen at a location where the NIL first
start to bend away to the north, forming a radio
lobe. 
\cite{kraft2008evidence} explain this bending 
as due to a pressure discontinuity in the gas.
\cite{1984gopal} had proposed that NIL is formed
by the jet interaction with the tidal structures.
The interaction between the jet and the gas 
cloud associated with the tidal structures 
might be triggering star formation in B and C.

\begin{figure}
	\includegraphics[width=\columnwidth]{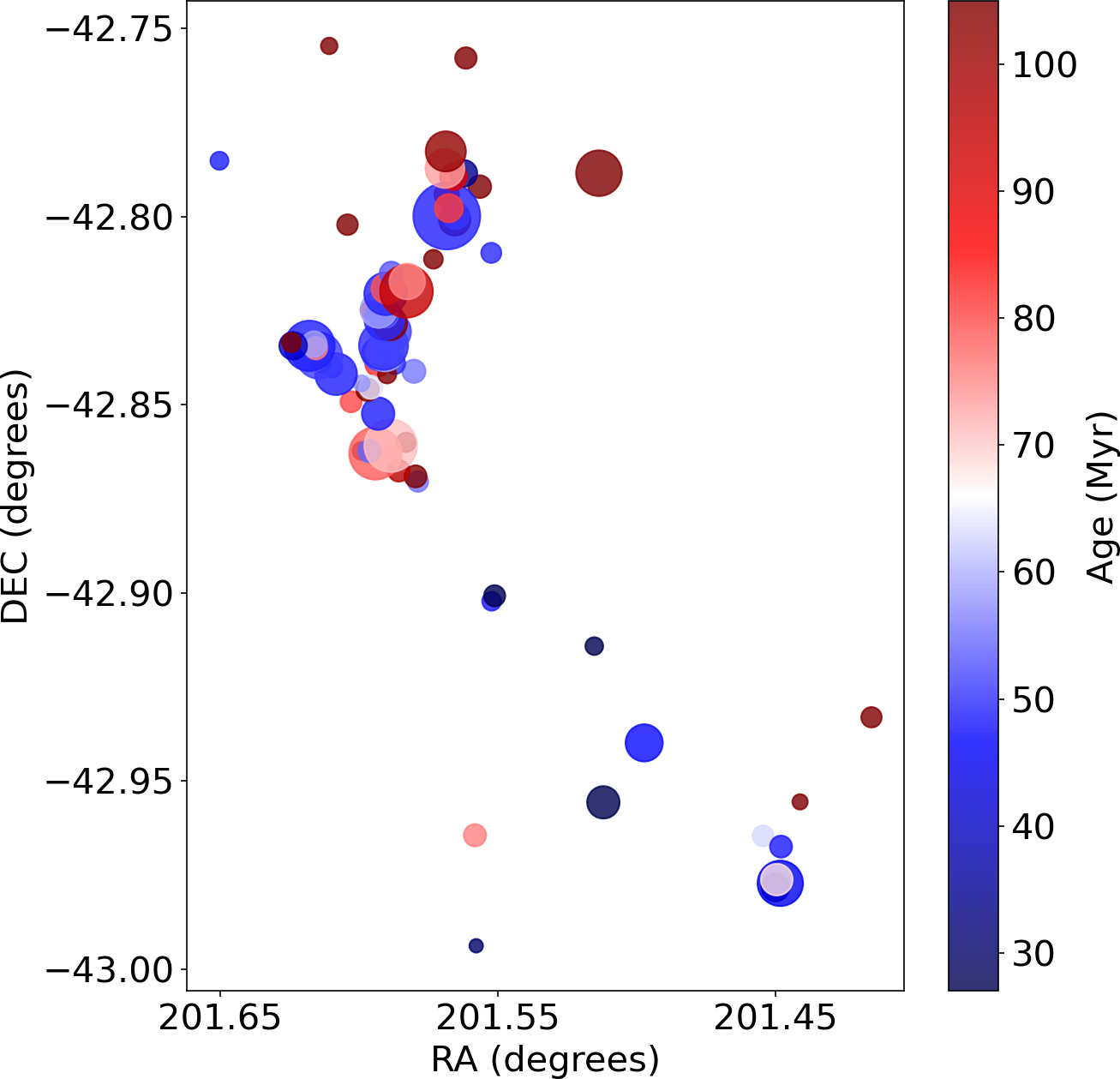}
    \caption{The age map of the NSR star-forming sources 
             from the Inner
             and Outer regions is shown, with 
             the size of circles being proportional
             to the isophotal area of sources.
             }
    \label{fig:age_map}
\end{figure}

Our age estimates in the Outer region imply that
most of the star formation traced in UV took place 
<100 Myr ago.
\cite{rejkuba2002radio} found 
that the youngest stars in the outer filament could have 
an age of 10 Myr for an abundance of Z = 0.004.
The Inner region sources are relatively young
(see Fig.~\ref{fig:age_distribution}).
Fig.~\ref{fig:age_map} shows an on-sky 
distribution of source ages, with the size of 
circles being proportional to the isophotal 
area of the sources.
 \cite{crockett2012triggered} derived an age 
of 1--4 Myr using HST observations for the 
southwest tip of the Inner filament. 
Our age estimate for the same sources lies 
at 3--4 Myr.

Our age estimates could be affected by 
contamination from foreground 
stars in the NUV bands, which may lead to 
older estimated ages for some sources.
We are also taking a single colour excess value
for each region.
Although the inner age distribution is sparse,
if the differences between the two distributions
are real, then it suggests that the Inner region 
contains stellar populations formed at a different
epoch than the Outer region.
The colour excess estimation is crucial 
in determining the differences between the two 
age distributions.
Suppose we increase the colour excess value 
of the Outer region to 0.15 and keep the 
Inner region colour excess value at its 
estimate given in Table~\ref{tab:age_summary}. 
In that case, the median ages of both 
regions will match.
Similarly, a match between median ages can be 
achieved if we lower the Inner region colour 
excess value to 0.14 and keep the Outer region 
colour excess unchanged.
Therefore, we caution that colour excess 
estimate offsets can also explain the age 
difference.

The age of Cen A giant radio lobes is estimated
to be $\sim$1 Gyr \citep{eilek2014dynamicmodelRadio}. 
Our median age estimate for the star-forming sources
in the Outer region is \outermedian{} Myr. 
The AGN activity is older than the star-forming
sources. 
Therefore, if past AGN jet activity is responsible 
for the observed star formation in the NML, the gas 
cloud from which the star formation is taking place may have 
come into contact with the jet around \outermedian{} 
Myr ago.
While this jet is currently not observed, 
it may have been present in the past. 
Presently, the NML is known to
harbour radio \& X-ray knots, ribbon-like FUV 
emission, star-forming sources, and molecular clouds. 
In our 2-dimensional view of NML, both 
molecular clouds and a part of the FUV ribbon 
occupy the same location in the sky.
The ribbon-like diffuse FUV emission could be
from gas heated by energy injection;
if molecular clouds are in the same region, 
they should also be heated up, and we 
should not be able to observe any molecular 
clouds.
However, \cite{salome2016atomic} noted high 
velocity dispersion for clouds in this region.
They suggested that this could
be either due to turbulence or several clouds
being in one location. 
If the high velocity dispersion is due to turbulence,
these clouds could have been partially affected 
by the energy injection from the jet that ionised the gas.

It should be kept in mind that the 
source distribution is 3-dimensional; 
therefore, what appears to be near one another
could be far apart. 
We consider the following to address the 
3-dimensional scenario in the Outer region.
The radio/X-ray knots 
and FUV ribbon should be close to the 
jet axis than the molecular clouds and 
star-forming sources.
The star formation may not occur if the imparted
energy is high; therefore, we do not expect to see it
near the radio \& X-ray knots where large amounts of 
energy may be injected by the jet. 
Star formation may get triggered at locations away
from the jet where the energy levels become moderate. 
Even further away from the jet axis, the imparted 
energies become substantially low so that it
does not trigger any more star formation.

If we assume that the present jet activity
is behind the recent star formation in the 
Inner region,
it is interesting to consider the observed difference in ages 
and the changes in AGN activity that lead to 
different histories in observed sources.
\cite{saikia2009recurrent} argue that AGN 
activity could be episodic, using Cen A as 
one among their examples.
Another possibility is a precessing jet; 
presently, the jet could be being deflected near 
the NIL, as \cite{1984gopal} explained, 
but it could have been aligned along the NML 
vector path in the past.
For the Inner region, the age distribution is 
comparatively more spread out and generally 
younger than in the Outer region.

\begin{table*}
\caption{Summary of selected sources with jet induced star formation from literature.}
\label{tab:jetinduced}
\begin{tabular}{@{}lrrll@{}}
\toprule
\multicolumn{1}{c}{Name}               & \multicolumn{1}{c}{Redshift} & \multicolumn{1}{c}{Distance to prominent structure} & \multicolumn{1}{c}{Morphology}                                & \multicolumn{1}{c}{Recent merger history}                                                   \\ \midrule
Cen A              & 0.001826                     & $\sim$15 kpc                                        & bifurcated (this study)                                       & yes \citep{wang2020mergerage}                                                        \\
MO & 0.0181                       & $\sim$20 kpc \citep{salome2015jet}                  & bifurcated \citep{croft2006minkowski}                         & yes (cluster level) \citep{bogdan2011chandra} \\
3C 285             & 0.0794                       & $\sim$70 kpc \citep{salome2015jet}                  & likely bifurcated \citep{van1993induced3c285} & yes \citep{salome2015jet}                                                            \\
3C 34              & 0.689                        & $\sim$120 kpc \citep{best1997jet}                   & bifurcated \citep{best1997jet}                                & unknown                                                                              \\
4C 41.17           & 3.8                          & $\sim$10 kpc \citep{bicknell2000jet4c41}                & bifurcated \citep{bicknell2000jet4c41}                            & yes \citep{de2005detection}                                                          \\ \bottomrule
\end{tabular}

\end{table*}

\section{Summary}

We observed the NSR of Cen A in both FUV 
and NUV using UVIT. 
We summarise our results below.

1. Dividing the NSR into Outer and Inner regions, 
we identified 61 star-forming sources in the 
Outer region and 16 sources in the Inner region.

2. The Outer region contains an extended 
structure of star formation oriented along 
the jet vector.
The structure shows a bifurcation, with one arm 
being more extended than the other arm. 
The Inner region contains less extended 
star-forming structures.

3. By comparing the model calculated colours 
from starburst99 models with observed FUV-NUV 
colours, we found the age of the star-forming
sources in the Outer region range between 
\outerspread{} Myr with a median age of 
\outermedian{} Myr.
Similarly, in the Inner region, we found the 
age of the star-forming sources to vary between 
\innerspread{} Myr with a median age of 
\innermedian{} Myr.

4. We arrived at a catalogue of UV sources 
likely associated with Cen A. 
Our observations tend to support jet induced 
star formation in Cen A.



\section*{Acknowledgements}

We would like to thank N. Junkes for sharing the 
Parkes radio telescope observations of Cen A. 
Astropy, IPython, Matplotlib, NumPy, Photutils, 
SAOImage DS9, and TOPCAT were used for data analysis, 
viewing, and plotting 
\citep{astropy:2013, astropy:2018, ipython2007, matplotlib, 
       numpy, photutils, ds9, taylor2005topcat}.
This research has made use of the Aladin sky atlas, 
VizieR catalogue access tool, SIMBAD database, 
and cross-match service provided by CDS, Strasbourg, France. 
This publication uses the data from the AstroSat 
mission of the Indian Space Research  Organisation (ISRO), 
archived at the Indian Space Science Data Centre (ISSDC).
This publication uses UVIT data processed by 
the payload operations centre at IIA. 
The UVIT is built in collaboration between IIA, 
IUCAA, TIFR, ISRO and CSA.

\section*{Data Availability}

The AstroSat UVIT data analysed here is associated 
with the Observation ID: G08\_023T01\_9000001978. 
The data is publicly available from the ISSDC 
Astrobrowse archive website. 
The data and Python scripts used for the
analysis are available from the repository: 
 \url{https://github.com/prajwel/Centaurus-A_NSR}.


\bibliographystyle{mnras}
\bibliography{references} 

\begin{thebibliography}{}
\makeatletter
\relax
\def\mn@urlcharsother{\let\do\@makeother \do\$\do\&\do\#\do\^\do\_\do\%\do\~}
\def\mn@doi{\begingroup\mn@urlcharsother \@ifnextchar [ {\mn@doi@}
  {\mn@doi@[]}}
\def\mn@doi@[#1]#2{\def\@tempa{#1}\ifx\@tempa\@empty \href
  {http://dx.doi.org/#2} {doi:#2}\else \href {http://dx.doi.org/#2} {#1}\fi
  \endgroup}
\def\mn@eprint#1#2{\mn@eprint@#1:#2::\@nil}
\def\mn@eprint@arXiv#1{\href {http://arxiv.org/abs/#1} {{\tt arXiv:#1}}}
\def\mn@eprint@dblp#1{\href {http://dblp.uni-trier.de/rec/bibtex/#1.xml}
  {dblp:#1}}
\def\mn@eprint@#1:#2:#3:#4\@nil{\def\@tempa {#1}\def\@tempb {#2}\def\@tempc
  {#3}\ifx \@tempc \@empty \let \@tempc \@tempb \let \@tempb \@tempa \fi \ifx
  \@tempb \@empty \def\@tempb {arXiv}\fi \@ifundefined
  {mn@eprint@\@tempb}{\@tempb:\@tempc}{\expandafter \expandafter \csname
  mn@eprint@\@tempb\endcsname \expandafter{\@tempc}}}

\bibitem[\protect\citeauthoryear{{Astropy Collaboration} et~al.,}{{Astropy
  Collaboration} et~al.}{2013}]{astropy:2013}
{Astropy Collaboration} et~al., 2013, \mn@doi [\aap]
  {10.1051/0004-6361/201322068}, \href
  {http://adsabs.harvard.edu/abs/2013A%26A...558A..33A} {558, A33}

\bibitem[\protect\citeauthoryear{{Astropy Collaboration} et~al.,}{{Astropy
  Collaboration} et~al.}{2018}]{astropy:2018}
{Astropy Collaboration} et~al., 2018, \mn@doi [\aj] {10.3847/1538-3881/aabc4f},
  \href {https://ui.adsabs.harvard.edu/abs/2018AJ....156..123A} {156, 123}

\bibitem[\protect\citeauthoryear{Auld et~al.,}{Auld
  et~al.}{2012}]{auld2012herschel}
Auld R.,  et~al., 2012, Monthly Notices of the Royal Astronomical Society, 420,
  1882

\bibitem[\protect\citeauthoryear{Bailer-Jones, Rybizki, Fouesneau, Demleitner
  \& Andrae}{Bailer-Jones et~al.}{2021}]{bailer2021estimating}
Bailer-Jones C.,  Rybizki J.,  Fouesneau M.,  Demleitner M.,   Andrae R.,
  2021, The Astronomical Journal, 161, 147

\bibitem[\protect\citeauthoryear{Best, Longair  \& R{\"o}ttgering}{Best
  et~al.}{1997}]{best1997jet}
Best P.,  Longair M.,   R{\"o}ttgering H.,  1997, Monthly Notices of the Royal
  Astronomical Society, 286, 785

\bibitem[\protect\citeauthoryear{Bianchi}{Bianchi}{2011}]{bianchi2011galex}
Bianchi L.,  2011, Astrophysics and Space Science, 335, 51

\bibitem[\protect\citeauthoryear{Bicknell, Sutherland, Van~Breugel, Dopita, Dey
   \& Miley}{Bicknell et~al.}{2000}]{bicknell2000jet4c41}
Bicknell G.~V.,  Sutherland R.~S.,  Van~Breugel W.~J.,  Dopita M.~A.,  Dey A.,
   Miley G.~K.,  2000, The Astrophysical Journal, 540, 678

\bibitem[\protect\citeauthoryear{Bogd{\'a}n et~al.,}{Bogd{\'a}n
  et~al.}{2011}]{bogdan2011chandra}
Bogd{\'a}n {\'A}.,  et~al., 2011, The Astrophysical Journal, 743, 59

\bibitem[\protect\citeauthoryear{Bolton, Stanley  \& Slee}{Bolton
  et~al.}{1949}]{bolton1949positions}
Bolton J.,  Stanley G.,   Slee O.,  1949, Nature, 164, 101

\bibitem[\protect\citeauthoryear{Bradley et~al.,}{Bradley
  et~al.}{2021}]{photutils}
Bradley L.,  et~al., 2021, astropy/photutils: 1.2.0,
  \mn@doi{10.5281/zenodo.5525286}, \url
  {https://doi.org/10.5281/zenodo.5525286}

\bibitem[\protect\citeauthoryear{Calzetti, Armus, Bohlin, Kinney, Koornneef  \&
  Storchi-Bergmann}{Calzetti et~al.}{2000}]{calzetti2000dust}
Calzetti D.,  Armus L.,  Bohlin R.~C.,  Kinney A.~L.,  Koornneef J.,
  Storchi-Bergmann T.,  2000, The Astrophysical Journal, 533, 682

\bibitem[\protect\citeauthoryear{Canning, Fabian, Johnstone, Sanders,
  Conselice, Crawford, Gallagher~III  \& Zweibel}{Canning
  et~al.}{2010}]{ngc1275canning2010star}
Canning R.,  Fabian A.,  Johnstone R.,  Sanders J.,  Conselice C.,  Crawford
  C.,  Gallagher~III J.,   Zweibel E.,  2010, Monthly Notices of the Royal
  Astronomical Society, 405, 115

\bibitem[\protect\citeauthoryear{Canning et~al.,}{Canning
  et~al.}{2014}]{ngc1275canning2014filamentary}
Canning R.,  et~al., 2014, Monthly Notices of the Royal Astronomical Society,
  444, 336

\bibitem[\protect\citeauthoryear{Charmandaris, Combes  \& Van
  Der~Hulst}{Charmandaris et~al.}{2000}]{charmandaris2000moleculargas}
Charmandaris V.,  Combes F.,   Van Der~Hulst J.,  2000, arXiv preprint
  astro-ph/0003175

\bibitem[\protect\citeauthoryear{Crockett et~al.,}{Crockett
  et~al.}{2012}]{crockett2012triggered}
Crockett R.~M.,  et~al., 2012, Monthly Notices of the Royal Astronomical
  Society, 421, 1603

\bibitem[\protect\citeauthoryear{Croft et~al.,}{Croft
  et~al.}{2006}]{croft2006minkowski}
Croft S.,  et~al., 2006, The Astrophysical Journal, 647, 1040

\bibitem[\protect\citeauthoryear{De~Breuck, Downes, Neri, Van~Breugel, Reuland,
  Omont  \& Ivison}{De~Breuck et~al.}{2005}]{de2005detection}
De~Breuck C.,  Downes D.,  Neri R.,  Van~Breugel W.,  Reuland M.,  Omont A.,
  Ivison R.,  2005, Astronomy \& Astrophysics, 430, L1

\bibitem[\protect\citeauthoryear{Di~Matteo, Bournaud, Martig, Combes, Melchior
  \& Semelin}{Di~Matteo et~al.}{2008}]{di2008frequency}
Di~Matteo P.,  Bournaud F.,  Martig M.,  Combes F.,  Melchior A.-L.,   Semelin
  B.,  2008, Astronomy \& Astrophysics, 492, 31

\bibitem[\protect\citeauthoryear{D{\"o}bereiner, Junkes, Wagner, Zinnecker,
  Fosbury, Fabbiano  \& Schreier}{D{\"o}bereiner
  et~al.}{1996}]{dobereiner1996rosat}
D{\"o}bereiner S.,  Junkes N.,  Wagner S.,  Zinnecker H.,  Fosbury R.,
  Fabbiano G.,   Schreier E.,  1996, The Astrophysical Journal, 470, L15

\bibitem[\protect\citeauthoryear{Duc \& Renaud}{Duc \&
  Renaud}{2013}]{duc2013tidesbook}
Duc P.-A.,  Renaud F.,  2013, Tides in astronomy and astrophysics, pp 327--369

\bibitem[\protect\citeauthoryear{Dufour \& Van~den Bergh}{Dufour \& Van~den
  Bergh}{1978}]{dufour1978inner}
Dufour R.~J.,  Van~den Bergh S.,  1978, The Astrophysical Journal, 226, L73

\bibitem[\protect\citeauthoryear{Eilek}{Eilek}{2014}]{eilek2014dynamicmodelRadio}
Eilek J.~A.,  2014, New Journal of Physics, 16, 045001

\bibitem[\protect\citeauthoryear{Ellison, Catinella  \& Cortese}{Ellison
  et~al.}{2018}]{ellison2018enhancedNeutralgas}
Ellison S.~L.,  Catinella B.,   Cortese L.,  2018, Monthly Notices of the Royal
  Astronomical Society, 478, 3447

\bibitem[\protect\citeauthoryear{Ellison, Viswanathan, Patton, Bottrell,
  McConnachie, Gwyn  \& Cuillandre}{Ellison
  et~al.}{2019}]{ellison2019mergertriggerAGN}
Ellison S.~L.,  Viswanathan A.,  Patton D.~R.,  Bottrell C.,  McConnachie
  A.~W.,  Gwyn S.,   Cuillandre J.-C.,  2019, Monthly Notices of the Royal
  Astronomical Society, 487, 2491

\bibitem[\protect\citeauthoryear{Feigelson, Schreier, Delvaille, Giacconi,
  Grindlay  \& Lightman}{Feigelson et~al.}{1981}]{feigelson1981xray}
Feigelson E.,  Schreier E.,  Delvaille J.,  Giacconi R.,  Grindlay J.,
  Lightman A.,  1981, The Astrophysical Journal, 251, 31

\bibitem[\protect\citeauthoryear{Fitzpatrick}{Fitzpatrick}{1999}]{fitzpatrick1999correcting}
Fitzpatrick E.~L.,  1999, Publications of the Astronomical Society of the
  Pacific, 111, 63

\bibitem[\protect\citeauthoryear{Fruscione et~al.,}{Fruscione
  et~al.}{2006}]{fruscione2006ciao}
Fruscione A.,  et~al., 2006, in Observatory Operations: Strategies, Processes,
  and Systems. p. 62701V

\bibitem[\protect\citeauthoryear{Gaibler, Khochfar, Krause  \& Silk}{Gaibler
  et~al.}{2012}]{gaibler2012jet}
Gaibler V.,  Khochfar S.,  Krause M.,   Silk J.,  2012, Monthly Notices of the
  Royal Astronomical Society, 425, 438

\bibitem[\protect\citeauthoryear{George et~al.,}{George
  et~al.}{2018}]{ngc7252george2018uvit}
George K.,  et~al., 2018, Astronomy \& Astrophysics, 613, L9

\bibitem[\protect\citeauthoryear{Ghosh et~al.,}{Ghosh
  et~al.}{2021a}]{ghosh2021orbit}
Ghosh S.,  et~al., 2021a, Journal of Astrophysics and Astronomy, 42, 1

\bibitem[\protect\citeauthoryear{Ghosh, Tandon, Joseph, Devaraj, Shelat  \&
  Stalin}{Ghosh et~al.}{2021b}]{ghosh2021performance}
Ghosh S.,  Tandon S.,  Joseph P.,  Devaraj A.,  Shelat D.,   Stalin C.,  2021b,
  Journal of Astrophysics and Astronomy, 42, 1

\bibitem[\protect\citeauthoryear{Ghosh et~al.,}{Ghosh
  et~al.}{2022}]{ghosh2022automated}
Ghosh S.,  et~al., 2022, arXiv preprint arXiv:2203.07693

\bibitem[\protect\citeauthoryear{{Gopal-Krishna} \&
  {Saripalli}}{{Gopal-Krishna} \& {Saripalli}}{1984}]{1984gopal}
{Gopal-Krishna} {Saripalli} L.,  1984, \aap, \href
  {https://ui.adsabs.harvard.edu/abs/1984A&A...141...61G} {141, 61}

\bibitem[\protect\citeauthoryear{{Gopal-Krishna} \& {Wiita}}{{Gopal-Krishna} \&
  {Wiita}}{2010}]{2010gopal}
{Gopal-Krishna} {Wiita} P.~J.,  2010, \mn@doi [\na]
  {10.1016/j.newast.2009.06.001}, \href
  {https://ui.adsabs.harvard.edu/abs/2010NewA...15...96G} {15, 96}

\bibitem[\protect\citeauthoryear{Graham \& Fassett}{Graham \&
  Fassett}{2002}]{graham2002star}
Graham J.~A.,  Fassett C.~I.,  2002, The Astrophysical Journal, 575, 712

\bibitem[\protect\citeauthoryear{Hamer, Salom{\'e}, Combes  \&
  Salom{\'e}}{Hamer et~al.}{2015}]{hamer2015muse}
Hamer S.,  Salom{\'e} P.,  Combes F.,   Salom{\'e} Q.,  2015, Astronomy \&
  Astrophysics, 575, L3

\bibitem[\protect\citeauthoryear{Hardcastle, Worrall, Kraft, Forman, Jones  \&
  Murray}{Hardcastle et~al.}{2003}]{hardcastle2003radio}
Hardcastle M.,  Worrall D.,  Kraft R.,  Forman W.,  Jones C.,   Murray S.,
  2003, The Astrophysical Journal, 593, 169

\bibitem[\protect\citeauthoryear{Harris, Rejkuba  \& Harris}{Harris
  et~al.}{2010}]{harris2010}
Harris G.~L.,  Rejkuba M.,   Harris W.~E.,  2010, Publications of the
  Astronomical Society of Australia, 27, 457

\bibitem[\protect\citeauthoryear{Harris et~al.,}{Harris et~al.}{2020}]{numpy}
Harris C.~R.,  et~al., 2020, Nature, 585, 357

\bibitem[\protect\citeauthoryear{Hunter}{Hunter}{2007}]{matplotlib}
Hunter J.~D.,  2007, \mn@doi [Computing in Science \& Engineering]
  {10.1109/MCSE.2007.55}, 9, 90

\bibitem[\protect\citeauthoryear{Israel}{Israel}{1998}]{israel1998centaurus}
Israel F.,  1998, The Astronomy and Astrophysics Review, 8, 237

\bibitem[\protect\citeauthoryear{Joye \& Mandel}{Joye \& Mandel}{2003}]{ds9}
Joye W.~A.,  Mandel E.,  2003, in Astronomical data analysis software and
  systems XII. p.~489

\bibitem[\protect\citeauthoryear{Junkes, Haynes, Harnett  \& Jauncey}{Junkes
  et~al.}{1993}]{junkes1993radio}
Junkes N.,  Haynes R.,  Harnett J.,   Jauncey D.,  1993, Astronomy and
  Astrophysics, 269, 29

\bibitem[\protect\citeauthoryear{Kennicutt~Jr \& Evans}{Kennicutt~Jr \&
  Evans}{2012}]{kennicutt2012star}
Kennicutt~Jr R.~C.,  Evans N.~J.,  2012, Annual Review of Astronomy and
  Astrophysics, 50, 531

\bibitem[\protect\citeauthoryear{Kinman \& Brown}{Kinman \&
  Brown}{2014}]{kinman2014rrlyr}
Kinman T.,  Brown W.~R.,  2014, The Astronomical Journal, 148, 121

\bibitem[\protect\citeauthoryear{Komatsu et~al.,}{Komatsu
  et~al.}{2009}]{komatsu2009five}
Komatsu E.,  et~al., 2009, The Astrophysical Journal Supplement Series, 180,
  330

\bibitem[\protect\citeauthoryear{Kraft et~al.,}{Kraft
  et~al.}{2000}]{kraft2000chandra}
Kraft R.,  et~al., 2000, The Astrophysical Journal, 531, L9

\bibitem[\protect\citeauthoryear{Kraft, Forman, Jones, Murray, Hardcastle  \&
  Worrall}{Kraft et~al.}{2002}]{kraft2002chandra}
Kraft R.,  Forman W.,  Jones C.,  Murray S.,  Hardcastle M.,   Worrall D.,
  2002, The Astrophysical Journal, 569, 54

\bibitem[\protect\citeauthoryear{Kraft et~al.,}{Kraft
  et~al.}{2008}]{kraft2008evidence}
Kraft R.,  et~al., 2008, The Astrophysical Journal Letters, 677, L97

\bibitem[\protect\citeauthoryear{Kraft et~al.,}{Kraft
  et~al.}{2009}]{kraft2009jet}
Kraft R.,  et~al., 2009, The Astrophysical Journal, 698, 2036

\bibitem[\protect\citeauthoryear{Lang, Hogg, Mierle, Blanton  \& Roweis}{Lang
  et~al.}{2010}]{lang2010astrometry}
Lang D.,  Hogg D.~W.,  Mierle K.,  Blanton M.,   Roweis S.,  2010, The
  astronomical journal, 139, 1782

\bibitem[\protect\citeauthoryear{Leitherer et~al.,}{Leitherer
  et~al.}{1999}]{leitherer1999starburst99}
Leitherer C.,  et~al., 1999, The Astrophysical Journal Supplement Series, 123,
  3

\bibitem[\protect\citeauthoryear{Leitherer, Ot{\'a}lvaro, Bresolin, Kudritzki,
  Faro, Pauldrach, Pettini  \& Rix}{Leitherer
  et~al.}{2010}]{leitherer2010library}
Leitherer C.,  Ot{\'a}lvaro P. A.~O.,  Bresolin F.,  Kudritzki R.-P.,  Faro
  B.~L.,  Pauldrach A.~W.,  Pettini M.,   Rix S.~A.,  2010, The Astrophysical
  Journal Supplement Series, 189, 309

\bibitem[\protect\citeauthoryear{Leitherer, Ekstr{\"o}m, Meynet, Schaerer,
  Agienko  \& Levesque}{Leitherer et~al.}{2014}]{leitherer2014effects}
Leitherer C.,  Ekstr{\"o}m S.,  Meynet G.,  Schaerer D.,  Agienko K.~B.,
  Levesque E.~M.,  2014, The Astrophysical Journal Supplement Series, 212, 14

\bibitem[\protect\citeauthoryear{Malin, Quinn  \& Graham}{Malin
  et~al.}{1983}]{malin1983shell}
Malin D.,  Quinn P.,   Graham J.,  1983, The Astrophysical Journal, 272, L5

\bibitem[\protect\citeauthoryear{McCollough \& Rots}{McCollough \&
  Rots}{2005}]{mccollough2005impact}
McCollough M.,  Rots A.,  2005, in Astronomical Data Analysis Software and
  Systems XIV. p.~478

\bibitem[\protect\citeauthoryear{McKinley et~al.,}{McKinley
  et~al.}{2018}]{mckinley2018jet}
McKinley B.,  et~al., 2018, Monthly Notices of the Royal Astronomical Society,
  474, 4056

\bibitem[\protect\citeauthoryear{McKinley et~al.,}{McKinley
  et~al.}{2022}]{mckinley2022multi}
McKinley B.,  et~al., 2022, Nature Astronomy, 6, 109

\bibitem[\protect\citeauthoryear{Mondal, Subramaniam  \& George}{Mondal
  et~al.}{2018}]{mondal2018uvit}
Mondal C.,  Subramaniam A.,   George K.,  2018, The Astronomical Journal, 156,
  109

\bibitem[\protect\citeauthoryear{Moreno et~al.,}{Moreno
  et~al.}{2019}]{moreno2019mergersimulation}
Moreno J.,  et~al., 2019, Monthly Notices of the Royal Astronomical Society,
  485, 1320

\bibitem[\protect\citeauthoryear{Morganti, Killeen, Ekers  \&
  Oosterloo}{Morganti et~al.}{1999}]{morganti1999centaurus}
Morganti R.,  Killeen N.,  Ekers R.,   Oosterloo T.,  1999, Monthly Notices of
  the Royal Astronomical Society, 307, 750

\bibitem[\protect\citeauthoryear{Mould et~al.,}{Mould
  et~al.}{2000}]{mould2000jetCenA}
Mould J.~R.,  et~al., 2000, The Astrophysical Journal, 536, 266

\bibitem[\protect\citeauthoryear{Neff, Eilek  \& Owen}{Neff
  et~al.}{2015a}]{neff2015radio}
Neff S.~G.,  Eilek J.~A.,   Owen F.~N.,  2015a, The Astrophysical Journal, 802,
  87

\bibitem[\protect\citeauthoryear{Neff, Eilek  \& Owen}{Neff
  et~al.}{2015b}]{neff2015complex}
Neff S.~G.,  Eilek J.~A.,   Owen F.~N.,  2015b, The Astrophysical Journal, 802,
  88

\bibitem[\protect\citeauthoryear{Oosterloo \& Morganti}{Oosterloo \&
  Morganti}{2005}]{oosterloo2005anomalous}
Oosterloo T.~A.,  Morganti R.,  2005, Astronomy \& Astrophysics, 429, 469

\bibitem[\protect\citeauthoryear{Pan et~al.,}{Pan
  et~al.}{2018}]{pan2018enhanceMolecularGas}
Pan H.-A.,  et~al., 2018, The Astrophysical Journal, 868, 132

\bibitem[\protect\citeauthoryear{P\'erez \& Granger}{P\'erez \&
  Granger}{2007}]{ipython2007}
P\'erez F.,  Granger B.~E.,  2007, \mn@doi [Computing in Science and
  Engineering] {10.1109/MCSE.2007.53}, 9, 21

\bibitem[\protect\citeauthoryear{Quillen, Graham  \& Frogel}{Quillen
  et~al.}{1993}]{quillen1993warpeddisk}
Quillen A.,  Graham J.~R.,   Frogel J.~A.,  1993, The Astrophysical Journal,
  412, 550

\bibitem[\protect\citeauthoryear{Quillen, Brookes, Keene, Stern, Lawrence  \&
  Werner}{Quillen et~al.}{2006}]{quillen2006spitzer}
Quillen A.~C.,  Brookes M.~H.,  Keene J.,  Stern D.,  Lawrence C.~R.,   Werner
  M.~W.,  2006, The Astrophysical Journal, 645, 1092

\bibitem[\protect\citeauthoryear{Rejkuba, Minniti, Courbin  \& Silva}{Rejkuba
  et~al.}{2002}]{rejkuba2002radio}
Rejkuba M.,  Minniti D.,  Courbin F.,   Silva D.,  2002, The Astrophysical
  Journal, 564, 688

\bibitem[\protect\citeauthoryear{Robitaille, Deil  \& Ginsburg}{Robitaille
  et~al.}{2020}]{robitaille2020reproject}
Robitaille T.,  Deil C.,   Ginsburg A.,  2020, Astrophysics Source Code
  Library, pp ascl--2011

\bibitem[\protect\citeauthoryear{Saikia \& Jamrozy}{Saikia \&
  Jamrozy}{2009}]{saikia2009recurrent}
Saikia D.,  Jamrozy M.,  2009, Bull. Astr. Soc. India, 37, 63

\bibitem[\protect\citeauthoryear{Salom{\'e}, Salom{\'e}  \& Combes}{Salom{\'e}
  et~al.}{2015}]{salome2015jet}
Salom{\'e} Q.,  Salom{\'e} P.,   Combes F.,  2015, Astronomy \& Astrophysics,
  574, A34

\bibitem[\protect\citeauthoryear{Salom{\'e}, Salom{\'e}, Combes, Hamer  \&
  Heywood}{Salom{\'e} et~al.}{2016a}]{salome2016star}
Salom{\'e} Q.,  Salom{\'e} P.,  Combes F.,  Hamer S.,   Heywood I.,  2016a,
  Astronomy \& Astrophysics, 586, A45

\bibitem[\protect\citeauthoryear{Salom{\'e}, Salom{\'e}, Combes  \&
  Hamer}{Salom{\'e} et~al.}{2016b}]{salome2016atomic}
Salom{\'e} Q.,  Salom{\'e} P.,  Combes F.,   Hamer S.,  2016b, Astronomy \&
  Astrophysics, 595, A65

\bibitem[\protect\citeauthoryear{Salom{\'e}, Salom{\'e}, Miville-Desch{\^e}nes,
  Combes  \& Hamer}{Salom{\'e} et~al.}{2017}]{salome2017inefficient}
Salom{\'e} Q.,  Salom{\'e} P.,  Miville-Desch{\^e}nes M.-A.,  Combes F.,
  Hamer S.,  2017, Astronomy \& Astrophysics, 608, A98

\bibitem[\protect\citeauthoryear{Santoro, Oonk, Morganti, Oosterloo  \&
  Tadhunter}{Santoro et~al.}{2016}]{santoro2016embedded}
Santoro F.,  Oonk J.,  Morganti R.,  Oosterloo T.,   Tadhunter C.,  2016,
  Astronomy \& Astrophysics, 590, A37

\bibitem[\protect\citeauthoryear{Saxton, Sutherland  \& Bicknell}{Saxton
  et~al.}{2001}]{saxton2001centaurus}
Saxton C.~J.,  Sutherland R.~S.,   Bicknell G.~V.,  2001, The Astrophysical
  Journal, 563, 103

\bibitem[\protect\citeauthoryear{Schiminovich, Van~Gorkom, Van Der~Hulst  \&
  Kasow}{Schiminovich et~al.}{1994}]{schiminovich1994neutralhydrogen}
Schiminovich D.,  Van~Gorkom J.,  Van Der~Hulst J.,   Kasow S.,  1994, The
  Astrophysical Journal, 423, L101

\bibitem[\protect\citeauthoryear{Schlafly \& Finkbeiner}{Schlafly \&
  Finkbeiner}{2011}]{schlafly2011measuring}
Schlafly E.~F.,  Finkbeiner D.~P.,  2011, The Astrophysical Journal, 737, 103

\bibitem[\protect\citeauthoryear{Schlegel, Finkbeiner  \& Davis}{Schlegel
  et~al.}{1998}]{schlegel1998maps}
Schlegel D.~J.,  Finkbeiner D.~P.,   Davis M.,  1998, The Astrophysical
  Journal, 500, 525

\bibitem[\protect\citeauthoryear{Stetson}{Stetson}{1987}]{stetson1987daophot}
Stetson P.~B.,  1987, Publications of the Astronomical Society of the Pacific,
  99, 191

\bibitem[\protect\citeauthoryear{Taghizadeh-Popp et~al.,}{Taghizadeh-Popp
  et~al.}{2020}]{taghizadeh2020sciserver}
Taghizadeh-Popp M.,  et~al., 2020, Astronomy and Computing, 33, 100412

\bibitem[\protect\citeauthoryear{Tandon et~al.,}{Tandon
  et~al.}{2017a}]{tandon2017firstresults}
Tandon S.,  et~al., 2017a, Journal of Astrophysics and Astronomy, 38, 1

\bibitem[\protect\citeauthoryear{Tandon et~al.,}{Tandon
  et~al.}{2017b}]{tandon2017orbit}
Tandon S.,  et~al., 2017b, The Astronomical Journal, 154, 128

\bibitem[\protect\citeauthoryear{Tandon et~al.,}{Tandon
  et~al.}{2020}]{tandon2020additional}
Tandon S.,  et~al., 2020, The Astronomical Journal, 159, 158

\bibitem[\protect\citeauthoryear{Taylor}{Taylor}{2005}]{taylor2005topcat}
Taylor M.~B.,  2005, in Astronomical data analysis software and systems XIV.
  p.~29

\bibitem[\protect\citeauthoryear{Van~Breugel, Filippenko, Heckman  \&
  Miley}{Van~Breugel et~al.}{1985}]{van1985minkowski}
Van~Breugel W.,  Filippenko A.,  Heckman T.,   Miley G.,  1985, The
  Astrophysical Journal, 293, 83

\bibitem[\protect\citeauthoryear{V{\'a}zquez \& Leitherer}{V{\'a}zquez \&
  Leitherer}{2005}]{vazquez2005optimization}
V{\'a}zquez G.~A.,  Leitherer C.,  2005, The Astrophysical Journal, 621, 695

\bibitem[\protect\citeauthoryear{Wang, Hammer, Rejkuba, Crnojevi{\'c}  \&
  Yang}{Wang et~al.}{2020}]{wang2020mergerage}
Wang J.,  Hammer F.,  Rejkuba M.,  Crnojevi{\'c} D.,   Yang Y.,  2020, Monthly
  Notices of the Royal Astronomical Society, 498, 2766

\bibitem[\protect\citeauthoryear{Wilkins, Bunker, Stanway, Lorenzoni  \&
  Caruana}{Wilkins et~al.}{2011}]{wilkins2011uv}
Wilkins S.~M.,  Bunker A.~J.,  Stanway E.,  Lorenzoni S.,   Caruana J.,  2011,
  Monthly Notices of the Royal Astronomical Society, 417, 717

\bibitem[\protect\citeauthoryear{van Breugel \& Dey}{van Breugel \&
  Dey}{1993}]{van1993induced3c285}
van Breugel W.~J.,  Dey A.,  1993, The Astrophysical Journal, 414, 563

\makeatother
\end{thebibliography}



\appendix

\section{Colour excess estimation}
\label{sec:colour_excess}

\begin{figure}
	\includegraphics[width=\columnwidth]{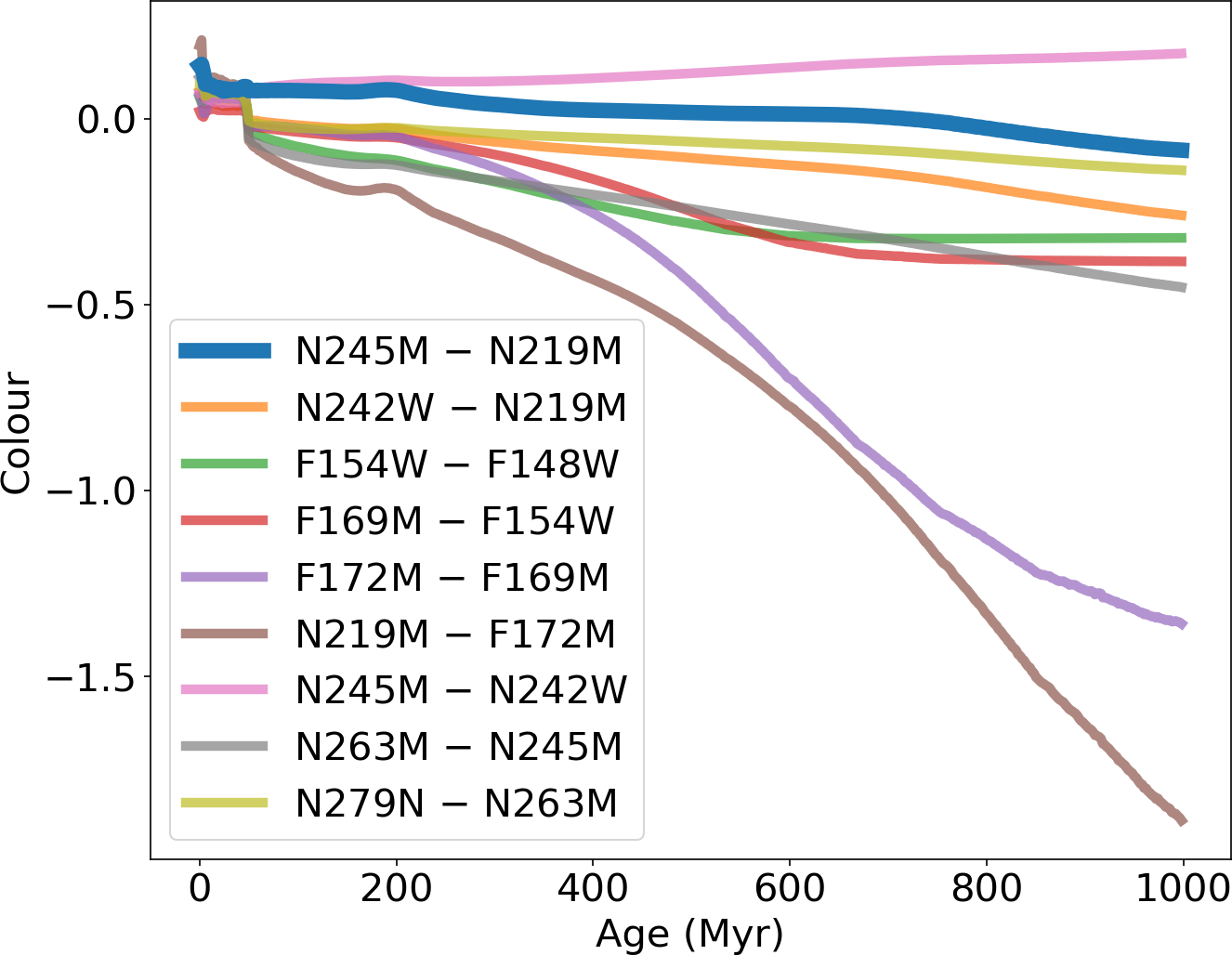}
    \caption{The plot shows simulated colours 
             with Starburst99 using v00-h for a sample of UVIT 
             filter combinations.
            }
    \label{fig:all_filter_colours}
\end{figure}

\begin{figure}
	\includegraphics[width=\columnwidth]{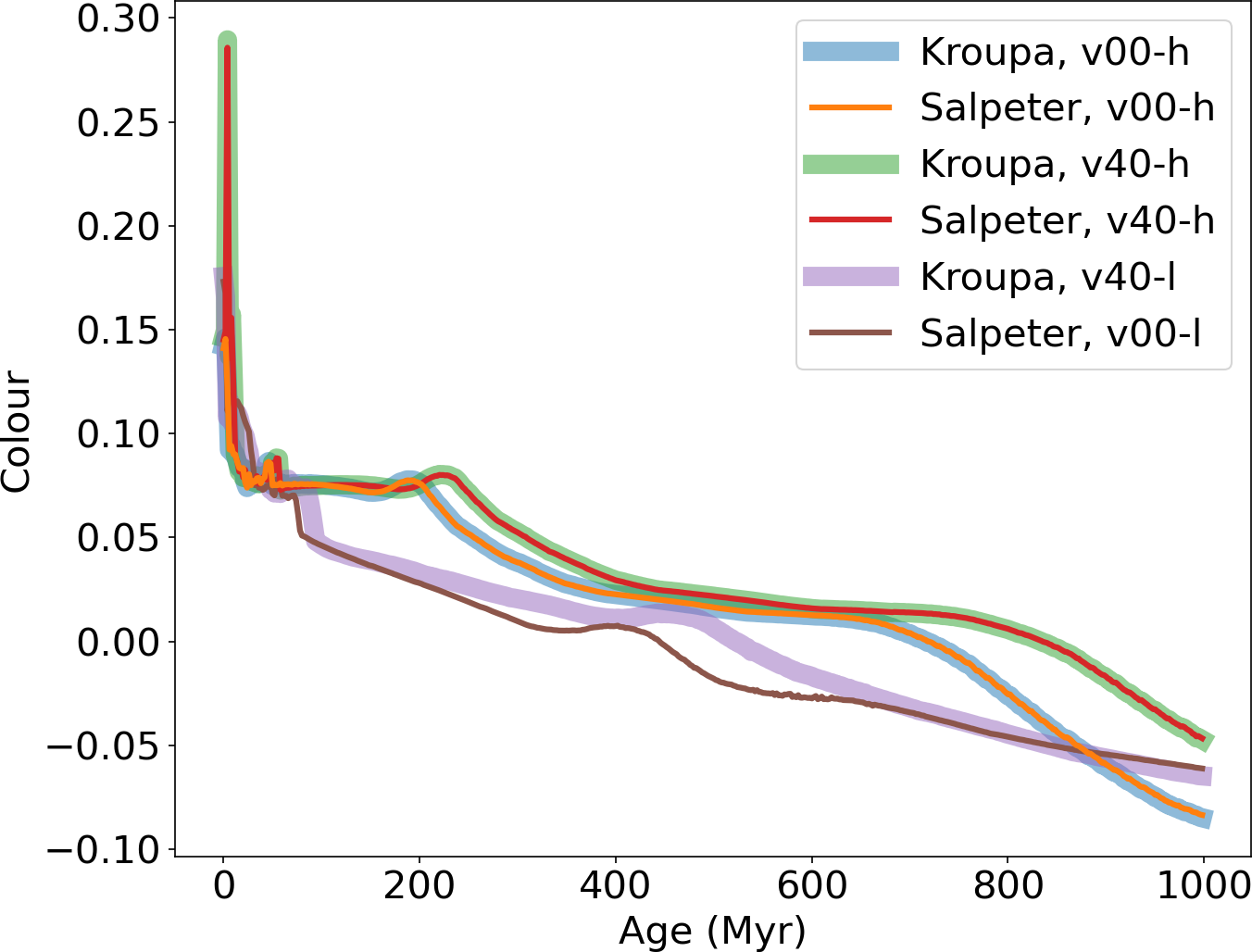}
    \caption{The plot shows N245M $-$ N219M colour
             generated from different simulations. 
             Kroupa and Salpeter IMFs were used in combinations. 
             The 0\% rotation model with solar metallicity 
             is denoted as v00-h and the 40\% model 
             as v40-h, while the 0\% rotation
             model with sub-solar metallicity 
             is mentioned as v00-l and the 40\% model as v40-l
             \citep{leitherer2014effects}.}
    \label{fig:colour_diff_simulations}
\end{figure}

We describe our approach to estimate the 
colour excess in the Outer and Inner regions.
It was observed that the simulated N245M $-$ N219M colour
remains approximately constant.
Fig.~\ref{fig:all_filter_colours} shows a 
plot of simulated colours
for a sample of UVIT filter combinations. 
A closer look (see Fig.~\ref{fig:colour_diff_simulations})
at the behaviour of N245M $-$ N219M colour
using simulations with different parameters 
reveals the following; 
a) Kroupa and Salpeter curves do not differ 
when other parameters are the same,
b) large variations occur up to 24 Myr,
c) there is a general downward trend as the age increases, and
d) lowering metallicity increases the downward slope of the curve. 
We do not consider the section below the 24 Myr age
for further calculations 
(this is also justified considering the ages derived from
the UV emitting population).
The main advantage of modelling the colour/slope of 
the medium band NUV filters to correct dust 
attenuation is that the same instrument data can be 
used.
For a detailed discussion on the behaviour of 
UV continuum slope, please see \cite{wilkins2011uv}.

\begin{figure}
	\includegraphics[width=\columnwidth]{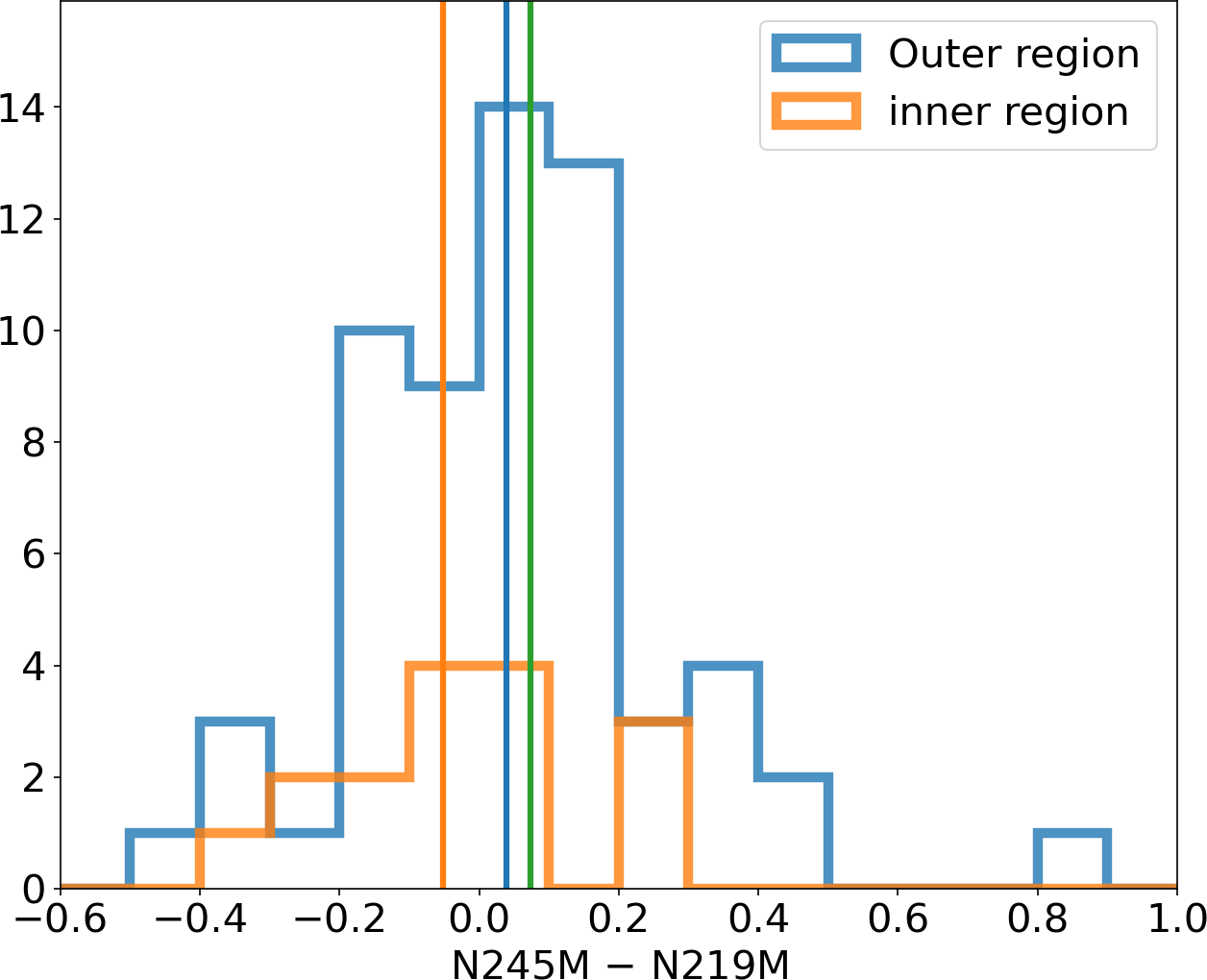}
    \caption{The plot shows the distribution of 
            observed N245M $-$ N219M colours,
             assuming zero attenuation in both regions. 
             The vertical blue and orange lines represent the 
             median values of the Outer and Inner region 
             distributions, respectively.
             The green vertical line represents the 
             0.074 colour.}
    \label{fig:without_attenuation_correction}
\end{figure}

\begin{figure}
	\includegraphics[width=\columnwidth]{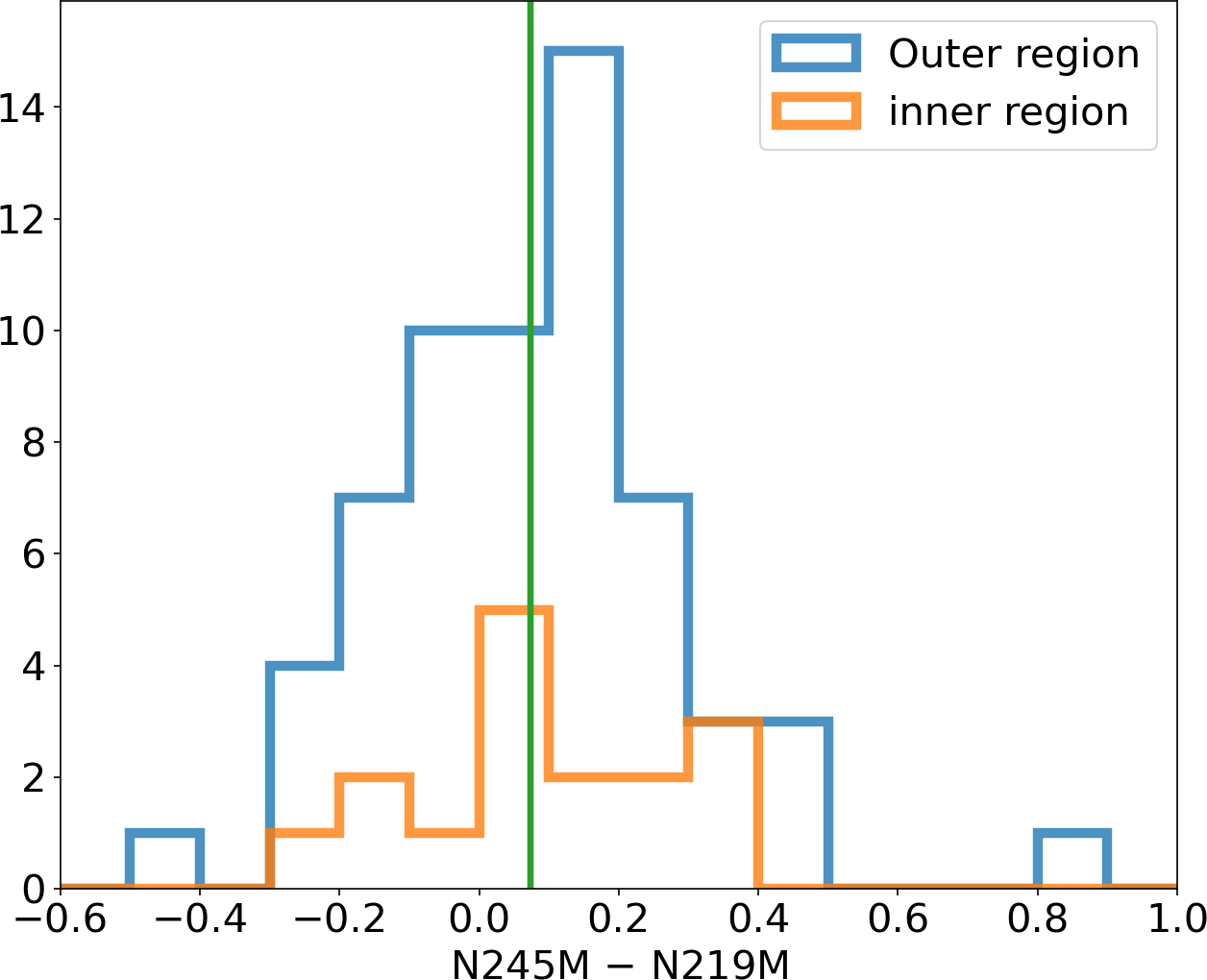}
    \caption{The plot shows the distribution of 
             observed N245M $-$ N219M colours
             after correcting for dust attenuation. 
             The green vertical line represents 
             the 0.074 colour.
             As in \ref{fig:without_attenuation_correction}, the vertical lines
             representing the median values of Outer and Inner region
             distributions have been plotted, but they are not visible
             because the green line covers them.
             }
    \label{fig:with_attenuation_correction}
\end{figure}

We used the v00-h model (solar metallicity, 0\% rotation) 
for our analysis.
For this model, 
the simulated N245M $-$ N219M colour has a median value 
of 0.074 in the range of 24--300 Myr, and the ratio 
becomes 0.014 in the range of 300--800 Myr
(see Fig.~\ref{fig:colour_diff_simulations}).
The distribution of the observed N245M $-$ N219M colours
is shown in Fig.~\ref{fig:without_attenuation_correction}, 
assuming zero dust extinction for the Inner 
and Outer regions. 
The median peak of the distributions is off 
from the expected value, and the Inner region 
appears to have a relatively
higher dust attenuation. 
We estimated the colour excess values for 
the Outer and Inner regions by comparing 
the simulated and observed colours
after assuming an attenuation law with $R_v = 4.05$
from \cite{calzetti2000dust}.
For both regions, we kept the expected colour as 0.074
since the median value of the region ages is estimated 
to be <300 Myr.
Thus estimated colour excess values are shown in 
Table~\ref{tab:age_summary}.
Fig.~\ref{fig:with_attenuation_correction} 
shows the distribution of N245M $-$ N219M 
colours after correcting for dust attenuation.

Instead of measuring a single colour 
excess value for each region, 
measuring them for each source was a 
tantalising prospect. 
We did not follow this due to the possible 
individual variations of the stellar 
populations from the employed 
model of burst star formation.
For example, ongoing/recent star formation can 
increase the observed N245M $-$ N219M colour.
This can be noticed in 
Fig.~\ref{fig:without_attenuation_correction};
0.074 is the expected colour value, 
while the observed distribution contains 
values above this threshold.  


\section{Tables}
\label{sec:age_tables}

The colours and estimated ages for the 
Outer and Inner regions are given in Tables~\ref{tab:inner_region_catalog}
\& \ref{tab:outer_region_catalog}, respectively.
For two sources in the Inner region, 
the observed colour values fell outside 
the simulated colour value range, so ages 
could not be determined. 
These two sources have a "nan" entry in the  
age column.

\begin{landscape}

\begin{table}
\centering
\caption{A sample table from the NSR inner region source catalogue and their properties is shown. 
RA and DEC are in degrees, the isophotal area of NSR sources are in pc$^2$, and flux values are in erg cm$^{-2}$ s$^{-1}$ \AA$^{-1}$.
The colour values used in the analysis are shown in columns
2, 3, and 4 from the end. Age is in Myr.
The whole catalogue can be accessed as a FITS table at
\url{https://github.com/prajwel/Centaurus-A_NSR/blob/main/cena_inr.fit}.}
\label{tab:inner_region_catalog}
\resizebox{\columnwidth}{!}{%
\begin{tabular}{@{}rrrrrrrrrrrrrrr@{}}
\toprule
\multicolumn{1}{c}{RA\_J2000} &
  \multicolumn{1}{c}{DEC\_J2000} &
  \multicolumn{1}{c}{Area} &
  \multicolumn{1}{c}{\begin{tabular}[c]{@{}c@{}}F148W\\ flux\end{tabular}} &
  \multicolumn{1}{c}{\begin{tabular}[c]{@{}c@{}}F148W\\ flux error\end{tabular}} &
  \multicolumn{1}{c}{\begin{tabular}[c]{@{}c@{}}N219M\\ flux\end{tabular}} &
  \multicolumn{1}{c}{\begin{tabular}[c]{@{}c@{}}N219M\\ flux error\end{tabular}} &
  \multicolumn{1}{c}{\begin{tabular}[c]{@{}c@{}}N245M\\ flux\end{tabular}} &
  \multicolumn{1}{c}{\begin{tabular}[c]{@{}c@{}}N245M\\ flux error\end{tabular}} &
  \multicolumn{1}{c}{\begin{tabular}[c]{@{}c@{}}N279N\\ flux\end{tabular}} &
  \multicolumn{1}{c}{\begin{tabular}[c]{@{}c@{}}N279N\\ flux error\end{tabular}} &
  \multicolumn{1}{c}{\begin{tabular}[c]{@{}c@{}}N219M -\\ F148W\end{tabular}} &
  \multicolumn{1}{c}{\begin{tabular}[c]{@{}c@{}}N245M -\\ F148W\end{tabular}} &
  \multicolumn{1}{c}{\begin{tabular}[c]{@{}c@{}}N245M -\\ N219M\end{tabular}} &
  \multicolumn{1}{c}{Age} \\ \midrule
201.4154820 &
  -42.9330740 &
  4.3644 &
  6.8203E-17 &
  4.5787E-18 &
  8.3704E-17 &
  9.4170E-18 &
  6.1774E-17 &
  5.9277E-18 &
  6.0927E-17 &
  7.2280E-18 &
  -0.5982 &
  -0.3773 &
  0.2209 &
  236.6650 \\
201.4412119 &
  -42.9555594 &
  2.4771 &
  3.4109E-17 &
  3.2836E-18 &
  4.0350E-17 &
  6.4850E-18 &
  4.1830E-17 &
  5.2128E-18 &
  6.1307E-17 &
  7.2152E-18 &
  -0.5583 &
  -0.7063 &
  -0.1480 &
  276.9896 \\
201.4480639 &
  -42.9674604 &
  5.0721 &
  6.4036E-17 &
  4.5647E-18 &
  4.3077E-17 &
  6.9508E-18 &
  3.6957E-17 &
  4.4060E-18 &
  2.5554E-17 &
  5.7344E-18 &
  0.0546 &
  0.1121 &
  0.0575 &
  48.5387 \\
201.4483249 &
  -42.9772165 &
  21.6451 &
  4.2856E-16 &
  1.1280E-17 &
  2.5638E-16 &
  1.6911E-17 &
  2.1368E-16 &
  1.1057E-17 &
  1.5934E-16 &
  1.3038E-17 &
  0.1819 &
  0.2709 &
  0.0889 &
  44.9612 \\
201.4495844 &
  -42.9762409 &
  10.4982 &
  2.1572E-16 &
  7.9684E-18 &
  1.9138E-16 &
  1.4552E-17 &
  1.4335E-16 &
  9.4402E-18 &
  1.4987E-16 &
  1.2019E-17 &
  -0.2459 &
  -0.0410 &
  0.2048 &
  68.0601 \\
201.4501058 &
  -42.9783131 &
  8.1390 &
  1.1607E-16 &
  6.0756E-18 &
  5.4060E-17 &
  8.1737E-18 &
  5.9365E-17 &
  5.8119E-18 &
  3.6778E-17 &
  7.3107E-18 &
  0.4538 &
  0.2432 &
  -0.2105 &
  6.2405 \\
... &
  ... &
  ... &
  ... &
  ... &
  ... &
  ... &
  ... &
  ... &
  ... &
  ... &
  ... &
  ... &
  ... &
  ... \\
201.5140402 &
  -42.9548405 &
  32.5561 &
  1.1006E-15 &
  1.7209E-17 &
  5.2118E-16 &
  2.3844E-17 &
  4.4654E-16 &
  1.6551E-17 &
  3.1603E-16 &
  1.6565E-17 &
  0.4357 &
  0.4946 &
  0.0589 &
  nan \\
201.5153409 &
  -42.9141509 &
  3.3028 &
  4.6056E-17 &
  3.7783E-18 &
  2.5608E-17 &
  5.3880E-18 &
  1.6732E-17 &
  3.0136E-18 &
  1.7467E-17 &
  4.1789E-18 &
  0.2614 &
  0.6145 &
  0.3531 &
  12.0083 \\
201.5512368 &
  -42.9007761 &
  4.7183 &
  5.9382E-17 &
  4.3458E-18 &
  3.7225E-17 &
  6.7104E-18 &
  3.3646E-17 &
  4.1500E-18 &
  9.9782E-18 &
  3.4028E-18 &
  0.1312 &
  0.1320 &
  0.0009 &
  26.4789 \\
201.5523085 &
  -42.9022168 &
  3.7156 &
  4.9372E-17 &
  3.9368E-18 &
  3.1780E-17 &
  6.0076E-18 &
  2.8903E-17 &
  3.7015E-18 &
  1.8776E-17 &
  4.1279E-18 &
  0.1025 &
  0.0966 &
  -0.0059 &
  48.3958 \\
201.5578779 &
  -42.9938295 &
  1.9463 &
  2.3509E-17 &
  2.7668E-18 &
  1.6702E-17 &
  4.1912E-18 &
  1.0703E-17 &
  2.1187E-18 &
  2.7618E-18 &
  2.1292E-18 &
  -0.0047 &
  0.3695 &
  0.3742 &
  29.8782 \\
201.5583015 &
  -42.9644371 &
  5.1901 &
  6.6281E-17 &
  4.5970E-18 &
  5.0235E-17 &
  7.3577E-18 &
  5.1406E-17 &
  5.5332E-18 &
  3.1067E-17 &
  5.0883E-18 &
  -0.0749 &
  -0.2088 &
  -0.1339 &
  75.6371 \\ \bottomrule
\end{tabular}%
}
\end{table}

\begin{table}
\centering
\caption{A sample table from the NSR outer region source catalogue and their properties is shown. 
RA and DEC are in degrees, the isophotal area of NSR sources are in pc$^2$, and flux values are in erg cm$^{-2}$ s$^{-1}$ \AA$^{-1}$.
The colour values used in the analysis are shown in columns
2, 3, and 4 from the end. Age is in Myr.
The whole catalogue can be accessed as a FITS table at
\url{https://github.com/prajwel/Centaurus-A_NSR/blob/main/cena_otr.fit}.}
\label{tab:outer_region_catalog}
\resizebox{\columnwidth}{!}{%
\begin{tabular}{@{}rrrrrrrrrrrrrrr@{}}
\toprule
\multicolumn{1}{c}{RA\_J2000} &
  \multicolumn{1}{c}{DEC\_J2000} &
  \multicolumn{1}{c}{Area} &
  \multicolumn{1}{c}{\begin{tabular}[c]{@{}c@{}}F148W\\ flux\end{tabular}} &
  \multicolumn{1}{c}{\begin{tabular}[c]{@{}c@{}}F148W\\ flux error\end{tabular}} &
  \multicolumn{1}{c}{\begin{tabular}[c]{@{}c@{}}N219M\\ flux\end{tabular}} &
  \multicolumn{1}{c}{\begin{tabular}[c]{@{}c@{}}N219M\\ flux error\end{tabular}} &
  \multicolumn{1}{c}{\begin{tabular}[c]{@{}c@{}}N245M\\ flux\end{tabular}} &
  \multicolumn{1}{c}{\begin{tabular}[c]{@{}c@{}}N245M\\ flux error\end{tabular}} &
  \multicolumn{1}{c}{\begin{tabular}[c]{@{}c@{}}N279N\\ flux\end{tabular}} &
  \multicolumn{1}{c}{\begin{tabular}[c]{@{}c@{}}N279N\\ flux error\end{tabular}} &
  \multicolumn{1}{c}{\begin{tabular}[c]{@{}c@{}}N219M -\\ F148W\end{tabular}} &
  \multicolumn{1}{c}{\begin{tabular}[c]{@{}c@{}}N245M -\\ F148W\end{tabular}} &
  \multicolumn{1}{c}{\begin{tabular}[c]{@{}c@{}}N245M -\\ N219M\end{tabular}} &
  \multicolumn{1}{c}{Age} \\ \midrule
 201.5136391 &
  -42.7884113 &
  21.8810 &
  2.6128E-16 &
  9.1519E-18 &
  1.8580E-16 &
  1.4329E-17 &
  1.9783E-16 &
  9.6866E-18 &
  1.4483E-16 &
  1.1011E-17 &
  -0.3573 &
  -0.6264 &
  -0.2690 &
  209.5469 \\
201.5524421 &
  -42.8095976 &
  4.2464 &
  6.2863E-17 &
  4.3803E-18 &
  2.7491E-17 &
  5.5578E-18 &
  2.8642E-17 &
  3.9691E-18 &
  2.3407E-17 &
  4.8736E-18 &
  0.1706 &
  -0.0749 &
  -0.2455 &
  49.6602 \\
201.5565762 &
  -42.7919991 &
  5.4260 &
  9.4251E-17 &
  5.2759E-18 &
  6.6465E-17 &
  8.4315E-18 &
  4.9285E-17 &
  4.8603E-18 &
  3.7904E-17 &
  5.4073E-18 &
  -0.3482 &
  -0.2245 &
  0.1238 &
  119.8103 \\
201.5615720 &
  -42.7577931 &
  4.8362 &
  8.5044E-17 &
  5.0096E-18 &
  5.6723E-17 &
  7.7662E-18 &
  5.1956E-17 &
  5.0874E-18 &
  5.0952E-17 &
  6.4502E-18 &
  -0.2878 &
  -0.3934 &
  -0.1056 &
  159.8501 \\
201.5623241 &
  -42.7884567 &
  7.4903 &
  1.0291E-16 &
  5.6522E-18 &
  6.0106E-17 &
  7.8657E-18 &
  3.2898E-17 &
  3.9478E-18 &
  3.8995E-17 &
  6.1924E-18 &
  -0.1437 &
  0.3098 &
  0.4535 &
  34.1387 \\
201.5650921 &
  -42.7877506 &
  5.2491 &
  6.6719E-17 &
  4.5932E-18 &
  4.2052E-17 &
  7.1007E-18 &
  2.9404E-17 &
  3.4900E-18 &
  2.5523E-17 &
  4.9044E-18 &
  -0.2263 &
  -0.0388 &
  0.1875 &
  64.2013 \\
... &
  ... &
  ... &
  ... &
  ... &
  ... &
  ... &
  ... &
  ... &
  ... &
  ... &
  ... &
  ... &
  ... &
  ... \\
201.6163112 &
  -42.8338520 &
  6.9005 &
  1.3554E-16 &
  6.2641E-18 &
  8.3553E-17 &
  9.4318E-18 &
  4.9794E-17 &
  4.8141E-18 &
  4.2948E-17 &
  6.3404E-18 &
  -0.2022 &
  0.1589 &
  0.3610 &
  58.7741 \\
201.6174792 &
  -42.8360053 &
  16.1011 &
  4.3998E-16 &
  1.1007E-17 &
  2.6334E-16 &
  1.6534E-17 &
  1.9688E-16 &
  1.0224E-17 &
  1.4402E-16 &
  1.1316E-17 &
  -0.1702 &
  -0.0553 &
  0.1149 &
  54.3644 \\
201.6179312 &
  -42.8343234 &
  26.6583 &
  1.1427E-15 &
  1.7309E-17 &
  5.9292E-16 &
  2.4449E-17 &
  4.2543E-16 &
  1.5310E-17 &
  3.1180E-16 &
  1.6335E-17 &
  -0.0152 &
  0.1443 &
  0.1595 &
  48.7067 \\
201.6238190 &
  -42.8343045 &
  8.0211 &
  1.2099E-16 &
  6.0707E-18 &
  5.1107E-17 &
  7.5551E-18 &
  4.5637E-17 &
  4.8545E-18 &
  4.2354E-17 &
  6.0343E-18 &
  0.2082 &
  0.1302 &
  -0.0780 &
  40.4394 \\
201.6243912 &
  -42.8334176 &
  3.8336 &
  5.0124E-17 &
  3.9727E-18 &
  3.1639E-17 &
  5.6720E-18 &
  2.8135E-17 &
  3.8101E-18 &
  1.4648E-17 &
  3.6019E-18 &
  -0.2279 &
  -0.3014 &
  -0.0735 &
  112.0269 \\
201.6503655 &
  -42.7851075 &
  3.4207 &
  4.6086E-17 &
  3.7995E-18 &
  2.2889E-17 &
  5.3397E-18 &
  1.7770E-17 &
  2.8368E-18 &
  9.9784E-18 &
  3.2496E-18 &
  0.0324 &
  0.1063 &
  0.0739 &
  48.6584 \\ \bottomrule
\end{tabular}%
}
\end{table}

\end{landscape}


\bsp	
\label{lastpage}
\end{document}